\documentclass[a4paper,11pt]{article}
\usepackage[english]{babel}
\usepackage[utf8]{inputenc}
\usepackage{braket}
\usepackage{physics}
\usepackage{color}
\usepackage{amssymb}
\usepackage{amsmath}
\usepackage{graphicx}
\usepackage{epstopdf}
\usepackage{subcaption}
\usepackage{comment}
\usepackage{mathtools}
\usepackage{ragged2e}
\usepackage{soul}
\usepackage{hyperref}
\hypersetup{
	colorlinks=true,
	linkcolor=blue,
	filecolor=cyan, 
	urlcolor=blue,
	citecolor=red,
} 
\usepackage[titletoc,toc,title]{appendix}
\numberwithin{equation}{section}
\usepackage{cleveref}
\usepackage[margin=2.5cm]{geometry}
\usepackage{cite}
\usepackage{epsfig}
\usepackage{float}
\usepackage{cancel}
\usepackage{amsfonts}
\usepackage{enumitem}
\usepackage[font={footnotesize,it}]{caption}
\usepackage{authblk}

\begin{document}
	
	\title{Entanglement negativity in T$\overline{\text{T}}$-deformed CFT$_2$s}
	
	\author[]{Debarshi Basu \thanks{\noindent E-mail:~ {\tt debarshi@iitk.ac.in}}}
	\author[]{Lavish\thanks{\noindent E-mail:~ {\tt clavish@iitk.ac.in}}}
	\author[]{Boudhayan Paul\thanks{\noindent E-mail:~ {\tt paul@iitk.ac.in}}}

	\affil[]{
		Department of Physics,\\
		
		Indian Institute of Technology,\\ 
		
		Kanpur 208 016, India
	}

	\date{}
	
	\maketitle
	
	\thispagestyle{empty}
	
	\begin{abstract}
		
		\noindent
		We apply a suitable replica technique to develop a perturbative expression for the entanglement negativity of bipartite mixed states in T$\overline{\text{T}}$-deformed CFT$_2$s up to the first order in the deformation parameter. Utilizing our perturbative construction we compute the entanglement negativity for various bipartite mixed states involving two disjoint intervals, two adjacent intervals, and a single interval in a T$\overline{\text{T}}$-deformed CFT$_2$ at a finite temperature, in the large central charge limit. Subsequently, we advance appropriate holographic constructions to compute the entanglement negativity for such bipartite states in T$\overline{\text{T}}$-deformed thermal CFT$_2$s dual to BTZ black holes in a finite cut-off bulk geometry and find agreement with the corresponding field theoretic results in the limit of small deformation parameter.
		\justify

	\end{abstract}
	
	\clearpage
	
	\tableofcontents
	
	\clearpage

\section{Introduction}
Over the past few decades, diverse areas of physics ranging from quantum many body systems in condensed matter to quantum gravity and black holes, have seen tremendous progress with the toolbox of quantum entanglement. For bipartite pure states, the entanglement entropy, defined as the von-Neumann entropy of the corresponding reduced density matrix, correctly captures the entanglement structure. On the other hand, for bipartite mixed states or tripartite pure states, entanglement entropy fails to be a viable measure of the entanglement structure due to contributions from irrelevant classical and quantum correlations. To address this significant issue, various other measures for bipartite mixed state entanglement have been introduced in the literature. Among these, a computable entanglement monotone termed the entanglement negativity was introduced in the seminal work \cite{Vidal:2002zz}. This non-convex \cite{Plenio:2005cwa} entanglement measure serves as an upper bound on the distillable entanglement for a given mixed state.

Although the calculation of these entanglement measures in extended many body systems is in general computationally challenging, remarkably in $(1+1)$-dimensional conformal field theories (CFT$_2$s) the entanglement entropy for bipartite pure states may be explicitly computed utilizing a novel \textit{replica technique} \cite{Calabrese:2004eu,Calabrese:2009ez, Calabrese:2009qy}. Interestingly, a similar replica technique to compute the entanglement negativity for various bipartite mixed states in CFT$_2$s was introduced in \cite{Calabrese:2012ew,Calabrese:2012nk,Calabrese:2014yza}. 

With the advance of the holographic correspondence \cite{Maldacena:1997re}, there has been intense focus on the holographic characterization of the entanglement structure in conformal field theories with large central charge and a sparse spectrum which are dual to bulk Anti-de Sitter (AdS) geometries. Such advent was pioneered by the celebrated Ryu-Takayanagi formula \cite{Ryu:2006bv} which states that the entanglement entropy of a subsystem in a CFT$_d$ is given by the area of a co-dimension two minimal spacelike surface in the bulk dual AdS$_{d+1}$ geometry, homologous to the subsystem under consideration. Furthermore, a covariant generalization of this formula was proposed in \cite{Hubeny:2007xt}. These proposals were proved in a series of subsequent interesting communications \cite{Fursaev:2006ih,Casini:2011kv,Lewkowycz:2013nqa,Dong:2016hjy}. With these developments in characterizing the pure state entanglement, the authors in \cite{Rangamani:2014ywa, Chaturvedi:2016rft, Chaturvedi:2016rcn, Jain:2017aqk, Malvimat:2018txq, Chaturvedi:2017znc, Jain:2017uhe, Malvimat:2018ood,Basu:2022nds,Basu:2022reu} explored several holographic constructions\footnote{For analogues of these proposals in the context of flat holography, see \cite{Basu:2021axf}.} for characterizing the mixed state entanglement structure through the entanglement negativity\footnote{Note that, in \cite{Kudler-Flam:2018qjo, Kusuki:2019zsp, KumarBasak:2020eia, KumarBasak:2021lwm, Basu:2021awn}, an alternative holographic proposal based on the bulk entanglement wedge cross-section (EWCS) was also investigated.}, which reproduced the field theoretic results \cite{Kulaxizi:2014nma,Malvimat:2017yaj} in the large central charge limit. Interestingly, these geometric constructions were substantiated through the consideration of a replica-symmetry breaking saddle to the bulk gravitational path integral for the replica partition function in \cite{KumarBasak:2020ams,Dong:2021clv}.


In a separate context, Zamolodchikov showed in a seminal work \cite{Zamolodchikov:2004ce} that CFT$_2$s deformed by the determinant of the stress-energy tensor have a solvable structure in the sense that the energy spectrum and the partition function may be determined exactly. This particular class of \textit{irrelevant deformations} is generally termed as the T$\overline{\text{T}}$-deformations. The UV structure of such theories are non-local and there are an infinite number of possible RG flows to the same fixed point. Furthermore, a holographic dual for such theories which alter the UV physics must be different from asymptotically AdS geometries which correspond to UV fixed points i.e., CFTs. A particularly simple description for a holographic dual was provided in \cite{McGough:2016lol} which change the asymptotics of the dual AdS spacetime by putting a finite cut-off radius. This proposal has passed several tests including the matching of the bulk and boundary two-point function, the energy spectrum and the partition function \cite{McGough:2016lol}. For further developments in this direction, see \cite{Shyam:2017znq,Kraus:2018xrn,Cottrell:2018skz,Taylor:2018xcy,Hartman:2018tkw,Shyam:2018sro,Caputa:2019pam,Giveon:2017myj,Asrat:2017tzd}. The entanglement entropy for bipartite pure states in different T$\overline{\text{T}}$-deformed CFTs has been investigated in \cite{Donnelly:2018bef,Lewkowycz:2019xse,Chen:2018eqk,Banerjee:2019ewu,Jeong:2019ylz,Murdia:2019fax,Park:2018snf,Asrat:2019end,He:2019vzf,Grieninger:2019zts}. While the holographic entanglement entropy may be exactly computed via the Ryu-Takayanagi formula, for the field theoretic computations one needs to resort to conformal perturbation theory \cite{Jeong:2019ylz}. Furthermore, in \cite{Asrat:2020uib}, a total correlation measure for bipartite mixed states known as the reflected entropy \cite{Dutta:2019gen,Jeong:2019xdr} and the corresponding holographic dual, namely the minimal entanglement wedge cross-section \cite{Takayanagi:2017knl,Nguyen:2017yqw} were investigated.

The above developments bring into sharp focus the outstanding issue of chracterizing the mixed state entanglement structure in such T$\overline{\text{T}}$-deformed CFTs. In this article, we address this issue by studying the entanglement negativity for various bipartite mixed states in T$\overline{\text{T}}$-deformed CFT$_2$s. Motivated by \cite{Chen:2018eqk,Jeong:2019ylz}, we advance a suitable replica technique and subsequently a conformal perturbation theory for computing the entanglement negativity in T$\overline{\text{T}}$-deformed CFT$_2$s. Following this, we compute the entanglement negativity for two disjoint, two adjacent and a single interval in a thermal CFT$_2$ deformed by the T$\overline{\text{T}}$ operator. Furthermore, we utilize the holographic constructions in \cite{Chaturvedi:2016rcn, Jain:2017aqk,Malvimat:2018txq} to reproduce these field theoretic results in the large central charge limit. We would like to emphasize that the study of mixed state entanglement in T$\overline{\text{T}}$-deformed CFTs investigated in the present work provide interesting insights on the entanglement structure of UV non-local theories and information theoretic aspects of the RG group.

The rest of the article is organized as follows. In \cref{sec:review}, we review the basic features of T$\overline{\text{T}}$-deformed conformal field theories, the quantum information theoretic definition of the entanglement negativity and the corresponding replica technique in CFT$_2$s. Following this, in \cref{sec:EN}, we develop an appropriate replica technique to compute the entanglement negativity for various bipartite states in a CFT$_2$ with T$\overline{\text{T}}$-deformation. Utilizing this replica technique, we subsequently compute the entanglement negativity for the mixed state configurations of two disjoint, two adjacent and a single interval in a T$\overline{\text{T}}$-deformed CFT at a finite temperature defined on a temporally compactified cylinder. The holographic characterization of the entanglement negativity for such mixed states forms the subject matter of \cref{sec:HEN}. In \cref{app}, the technical details are collected. Finally, in \cref{sec:summary}, we provide a summary of our results and present a discussion of the future open issues.

\section{Review of earlier literature}
\label{sec:review}
\subsection{T$\overline{\text{T}}$-deformation in a CFT$_2$}
\label{sec:review1}
In this subsection we briefly review the salient features of two-dimensional conformal field theory deformed by the insertion of the following double-trace operator into the undeformed Lagrangian \cite{Zamolodchikov:2004ce}
\begin{align}
	\left<T\bar{T}\right>=\frac{1}{8}\left(\left<T_{ab}\right>\left<T^{ab}\right>-\left<T^a_a\right>^2\right)\,.
\end{align}
This composite operator, satisfying the factorization property \cite{Zamolodchikov:2004ce}, is called the T$\overline{\text{T}}$ operator and correspondingly the deformed CFT is termed a T$\overline{\text{T}}$-deformed CFT. The T$\overline{\text{T}}$ deformation of a CFT$_2$ defines a one parameter family of theories characterized by a deformation parameter $\mu\,(\geq 0)$ having the dimensions of length squared. The deformed theory is described by the flow equation \cite{Zamolodchikov:2004ce,Chen:2018eqk,Jeong:2019ylz}
\begin{align}
	\frac{d S_{\text{QFT}}^{(\mu)}}{d \mu} = \int d^{2} x\:(T\bar{T})_{\mu}~~, ~~ S_{\text{QFT}}^{(\mu)} \Bigg|_{\mu = 0} = S_{\text{CFT}}\, ,
\end{align}
where $S_{\text{QFT}}^{(\mu)}$ and $S_{\text{CFT}}$ are the actions of the deformed and undeformed theories respectively. The energy spectrum of a T$\overline{\text{T}}$-deformed CFT$_2$ is exactly solvable \cite{Smirnov:2016lqw,Cavaglia:2016oda}. 

Perturbatively, for a small deformation parameter $\mu$, the action of the deformed CFT may be written as \cite{Chen:2018eqk,Jeong:2019ylz}
\begin{align}
	S_{\text{QFT}}^{(\mu)}= S_{\text{CFT}}+\mu \int d^{2}x\;(T\bar{T})_{\mu=0}=S_{\text{CFT}}+\mu\int d^{2}x\;\left(T\bar{T}-\Theta^2\right)\, , \label{Def-action}
\end{align}
where $T\equiv T_{ww}$, $\bar{T}\equiv T_{\bar{w}\bar{w}}$ and $\Theta\equiv T_{w\bar{w}}$ are the components of the undeformed energy momentum tensor expressed in the complex coordinates $(w,\bar{w})$. In this manuscript, we always consider the deformed CFT on a cylinder for which the expectation value of $\Theta$ vanishes and hence we may omit the $\Theta^2$ term entirely \cite{Chen:2018eqk}.
A holographic description of T$\overline{\text{T}}$-deformed CFTs was given in \cite{McGough:2016lol}, for which the relevant discussions are deferred till \cref{sec:HEN}.
\subsection{Entanglement negativity in CFT$_2$s}
\label{sec:review2}
In this article, we will focus on a computable mixed state entanglement measure termed the entanglement negativity introduced in \cite{Vidal:2002zz}. This non-convex entanglement monotone\cite{Plenio:2005cwa} provides an upper bound to the distillable entanglement. For a bipartite mixed state $\rho_{AB}\in\mathcal{H}_A\otimes\mathcal{H}_B$, the logarithmic entanglement negativity between subsystems $A$ and $B$ is defined as the natural logarithm of the trace norm of the density matrix partially transposed with respect to the subsystem $B$ as 
\begin{align}
	\mathcal{E}(A:B):=\log||\rho_{AB}^{T_B}||\,,
\end{align}
where for an arbitrary hermitian matrix $M$ the trace norm is defined as $||M||=\text{Tr}\sqrt{M M^\dagger}$ and the partially transposed density matrix $\rho_{AB}^{T_B}$ is defined through the following operation
\begin{align}
	\big<i_A,j_B|\rho_{AB}^{T_B}|k_A,l_B\big>=\left<i_A,l_B|\rho_{AB}|k_A,j_B\right>\,,
\end{align}
with $\{i_A\}$ and $\{j_B\}$ representing orthogonal bases for the Hilbert spaces $\mathcal{H}_A$ and $\mathcal{H}_B$ respectively.

A replica technique to compute the entanglement negativity for bipartite states in a CFT$_2$ was developed in \cite{Calabrese:2012ew,Calabrese:2012nk,Calabrese:2014yza}, where one considers $n_e\in 2\mathbb{Z}^+$ copies of the original manifold $\mathcal{M}$, with branch cuts along the subsystems $A$ and $B$. Finally, the entanglement negativity for the bipartite state $\rho_{AB}$ is obtained by considering the even analytic continuation $n_e\to 1$ of the replica index as follows
\begin{align}
	\mathcal{E}(A:B)=\lim_{n_e\to 1}\mathcal{E}^{(n_e)}(A:B)\equiv\lim_{n_e\to 1}\log \Tr \left(\rho_{AB}^{T_B}\right)^{n_e}\,.\label{Neg-replica-def}
\end{align}
The Riemann surface computing the path integral for the trace in \cref{Neg-replica-def} is prepared via a particular gluing of the individual copies where the branch cuts along $A$ are sewed cyclically while those along $B$ are sewed anti-cyclically. The partition function on this replica manifold computes the Renyi entanglement negativity $\mathcal{E}^{(n_e)}$. Utilizing the replica technique, the entanglement negativity between two subsystems $A$ and $B$ in $\mathcal{M}$ may be expressed in terms of the logarithm of the (normalized) partition function on the $n_e$ sheeted Riemann surface as follows \cite{Calabrese:2012ew,Calabrese:2012nk,Calabrese:2014yza}
\begin{align}
	\mathcal{E}(A:B)=\lim_{n_e\to 1}\log\frac{\mathbb{Z}\left[\mathcal{M}_{n_e}\right]}{\left(\mathbb{Z}\left[\mathcal{M}\right]\right)^{n_e}}\,,\label{Neg-def}
\end{align}
where $\mathcal{M}_{n_e}$ denotes the $n_e$ sheeted Riemann surface glued cyclically along $A$ and anti-cyclically along $B$. In a CFT$_2$, the partition function in \cref{Neg-def} may be recast in terms of various correlation functions of \textit{twist operators} placed at the endpoints of the subsystems $A$ and $B$ in the orbifold theory $\tilde{\mathcal{M}}_{n_e}\equiv\mathcal{M}_{n_e}/Z_{n_e}$ obtained by quotienting via the replica $Z_{n_e}$ symmetry \cite{Calabrese:2012ew,Calabrese:2012nk,Calabrese:2014yza}. For example, in the case of two disjoint intervals $A=[z_1,z_2]$ and $B=[z_3,z_4]$ in the vacuum state of a CFT$_2$, the entanglement negativity between $A$ and $B$ may be expressed as \cite{Calabrese:2012ew,Calabrese:2012nk}
\begin{align}
	\mathcal{E}=\lim_{n_e\to 1}\log\left<\sigma_{n_e}(z_1)\bar{\sigma}_{n_e}(z_2)\bar{\sigma}_{n_e}(z_3)\sigma_{n_e}(z_4)\right>_{\tilde{\mathcal{M}}_{n_e}}\,,\label{Neg-dj-twist}
\end{align}
where $\sigma_{n_e}$ and $\bar{\sigma}_{n_e}$ are the twist and anti-twist fields respectively. These are primary operators in the CFT$_2$ with conformal dimensions
\begin{align}
	h_{n_e}=\bar{h}_{n_e}=\frac{c}{24}\left(n_e-\frac{1}{n_e}\right)\,.
\end{align}
\section{Entanglement negativity in T$\overline{\text{T}}$-deformed CFT$_2$}
\label{sec:EN}
In this section, we devise a suitable replica technique to compute the entanglement negativity for various bipartite mixed states in a CFT$_2$ perturbed by the T$\overline{\text{T}}$ operator. We utilize the twist operator formalism to compute the correlation function on the $n_e$-sheeted (with $n_e$ even) Riemann surface in the replica method.

Consider a T$\overline{\text{T}}$-deformed CFT$_2$ living on some manifold $\mathcal{M}$. We are concerned with calculating the entanglement negativity for bipartite mixed states consisting of two spatial intervals $A$ and $B$. The $n_e$-sheeted Riemann surface $\mathcal{M}_{n_e}$ is obtained by joining $n_e$ copies of the manifold $\mathcal{M}$, cyclically along $A$ and anti-cyclically along $B$. The partition function of the deformed theory may be written in the path integral representation as follows, 
\begin{align}
	\mathbb{Z}\left[\mathcal{M}_{n_e}\right] = \int_{\mathcal{M}_{n_e}} \mathcal{D}\phi\; e^{-S_\text{QFT}^{(\mu)}[\phi]}\,,
\end{align}
where $S_\text{QFT}^{(\mu)}$ is the action for the T$\overline{\text{T}}$-deformed CFT. For the case with a small deformation parameter in \cref{Def-action}, we may obtain the entanglement negativity from \cref{Neg-def} as,
\begin{align}
	\mathcal{E}^{(\mu)}(A:B)=\lim_{n_e\to 1}\log\left[\frac{\int_{\mathcal{M}_{n_e}} \mathcal{D}\phi\; e^{-S_\text{CFT}-\mu \int_{\mathcal{M}_{n_e}}(T\bar{T})}}{\left(\int_{\mathcal{M}} \mathcal{D}\phi\; e^{-S_\text{CFT}-\mu \int_{\mathcal{M}}(T\bar{T})}\right)^{n_e}}\right]\,,
\end{align}
where the superscript $\mu$ indicates that we are working with a deformed CFT$_2$. Since the deformation parameter $\mu$ is small, we may further expand the exponential in terms of $\mu$, to obtain
\begin{align}
	\mathcal{E}^{(\mu)}(A:B)&=\lim_{n_e\to 1}\log\left[\frac{\int_{\mathcal{M}_{n_e}} \mathcal{D}\phi\; e^{-S_\text{CFT}}\left(1-\mu \int_{\mathcal{M}_{n_e}}(T\bar{T})+\mathcal{O}(\mu^2)\right)}{\left[\int_{\mathcal{M}}\mathcal{D}\phi\; e^{-S_\text{CFT}}\left(1-\mu \int_{\mathcal{M}}(T\bar{T})+\mathcal{O}(\mu^2)\right)\right]^{n_e}}\right]\,,\\&= \mathcal{E}_{\text{CFT}}(A:B)+\lim_{n_e\to 1}\log\left[\frac{(1-\mu \int_{\mathcal{M}_{n_e}}\Braket{T\bar{T}}_{\mathcal{M}_{n_e}})}{\left(1-\mu\int_{\mathcal{M}}\Braket{T\bar{T}}_{\mathcal{M}}\right)^{n_e}}\right]\,.
\end{align}
Here $\mathcal{E}_{\text{CFT}}(A:B)$ is the entanglement negativity of the bipartite quantum state $\rho_{AB}$ in a $\text{CFT}_2$ and the expectation value of the T$\overline{\text{T}}$ operator on the manifold $\mathcal{M}$ is defined as (similarly on $\mathcal{M}_{n_e}$),
\begin{align}
	\Braket{T\bar{T}}_{\mathcal{M}} = \frac{\int_{\mathcal{M}} \mathcal{D}\phi\; e^{-S_\text{CFT}}(T\bar{T})}{\int_{\mathcal{M}} \mathcal{D}\phi\; e^{-S_\text{CFT}}}\,.
\end{align}
Therefore, the first order correction in the entanglement negativity of $\text{CFT}_2$ due to the deformation by the T$\overline{\text{T}}$ operator is given by
\begin{equation}
	\delta \mathcal{E}(A:B) = -\mu \lim_{n_e\to 1} \left[\int_{\mathcal{M}_{n_e}} \Braket{T\bar{T}}_{\mathcal{M}_{n_e}}-n_{e} \int_{\mathcal{M}}\Braket{T\bar{T}}_{\mathcal{M}}\right]\,.\label{varEN}
\end{equation}
In this article, we consider the deformed $\text{CFT}_{2}$ in an excited state at finite temperature $1/\beta$ and the manifold $\mathcal{M}$ is an infinitely long cylinder whose Euclidean time direction is compactified with the circumference $\beta$. We set up the complex coordinates $w=x+i \tau$ and $\bar{w}=x-i \tau$ on the cylinder $\mathcal{M}$, where $x \in (-\infty, \infty)$ and $\tau \in (0, \beta)$ with the periodic identification $\tau \sim \tau + \beta$. The cylinder is described by the following conformal map from a complex plane $\mathbb{C}$,
\begin{equation}
	z = e^{\frac{2 \pi w}{\beta}} ~~\,, ~~\bar{z} = e^{\frac{2 \pi \bar{w}}{\beta}}\,,\label{map}
\end{equation}
where $(z, \bar{z})$ are the coordinates on the complex plane.
Under this map, the energy momentum tensors transform as follows,
\begin{align}
	T(w) = T(z)-\frac{\pi^2 c}{6 \beta^2} ~~\,, ~~ \bar{T}(\bar{w}) = \bar{T} (\bar{z})-\frac{\pi^2 c}{6 \beta^2}\,.\label{Stress tensor}
\end{align}
Since for the vacuum state of a $\text{CFT}_2$ described at the complex plane, $\Braket{T(z)}_{\mathbb{C}}=\Braket{\bar{T}(\bar{z})}_{\mathbb{C}}=0$\,, we obtain
\begin{align}
	\Braket{T(w) \bar{T}(\bar{w})}_{\mathcal{M}} = \left(\frac{\pi^2 c}{6 \beta^2}\right)^2 \,. \label{tt-single sheet}
\end{align}

\subsection{Two disjoint intervals}
\label{subsec:disj-ft}
\begin{figure}[H]
	\centering
	\includegraphics[scale=0.6]{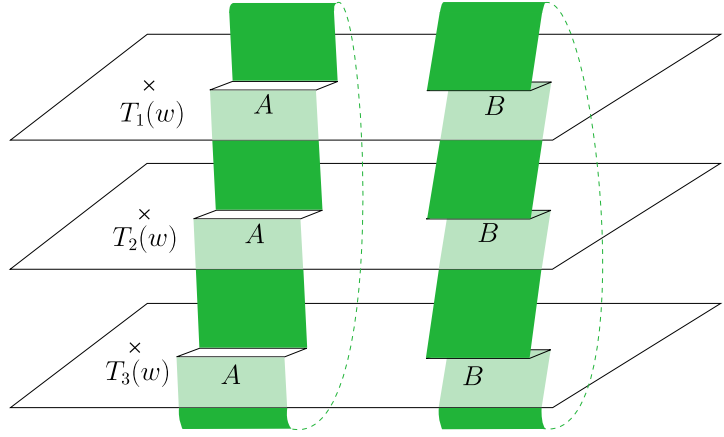}
	\caption{Schematics of the replica manifold computing the path integral for the entanglement negativity of two disjoint intervals $A$ and $B$ in a T$\overline{\text{T}}$-deformed CFT$_2$\,. Figure modified from \cite{Calabrese:2012nk}\,.}
	\label{fig:replica}
\end{figure}

In this subsection, we will calculate the first order correction in the entanglement negativity for a bipartite mixed state comprised of two disjoint spatial intervals $A=[x_1,x_2]$ and $B=[x_3,x_4]$ in a finite temperature T$\overline{\text{T}}$-deformed CFT$_2$. Consider a CFT$_2$ living on the cylindrical manifold $\mathcal{M}$ with temperature $1/\beta$, perturbed by T$\overline{\text{T}}$- deformation.
To compute the entanglement negativity in the mixed state $\rho_{AB}$ defined in a T$\overline{\text{T}}$-deformed CFT$_2$, we need to determine the expectation value of the T$\overline{\text{T}}$ operator. On the Riemann surface $\mathcal{M}_{n_e}$ (cf. \cref{fig:replica}), the value of $\Braket{T\bar{T}}_{\mathcal{M}_{n_e}}$ may be obtained from inserting the T$\overline{\text{T}}$ operator into the correlation function of the twist operators located at the end points of intervals $A$ and $B$ as follows \cite{Calabrese:2004eu,Calabrese:2009qy}
\begin{align}
	\int_{\mathcal{M}_{n_e}} \Braket{T\bar{T}}_{\mathcal{M}_{n_e}} &= \sum_{k=1}^{n_e} \int_{\mathcal{M}} \frac{\Braket{T_{k}(w)\bar{T}_{k}(\bar{w})\sigma_{n_e}(w_{1}, \bar{w}_{1})\bar{\sigma}_{n_e}(w_{2}, \bar{w}_{2})\bar{\sigma}_{n_e}(w_{3}, \bar{w}_{3})\sigma_{n_e}(w_{4}, \bar{w}_{4})}_\mathcal{M}}{\Braket{\sigma_{n_e}(w_{1}, \bar{w}_{1})\bar{\sigma}_{n_e}(w_{2}, \bar{w}_{2})\bar{\sigma}_{n_e}(w_{3}, \bar{w}_{3})\sigma_{n_e}(w_{4}, \bar{w}_{4})}_\mathcal{M}} \notag\\&= \int_{\mathcal{M}}\frac{1}{n_e} \frac{\Braket{T^{(n_e)}(w)\bar{T}^{(n_e)}(\bar{w})\sigma_{n_e}(w_{1}, \bar{w}_{1})\bar{\sigma}_{n_e}(w_{2}, \bar{w}_{2})\bar{\sigma}_{n_e}(w_{3}, \bar{w}_{3})\sigma_{n_e}(w_{4}, \bar{w}_{4})}_\mathcal{M}}{\Braket{\sigma_{n_e}(w_{1}, \bar{w}_{1})\bar{\sigma}_{n_e}(w_{2}, \bar{w}_{2})\bar{\sigma}_{n_e}(w_{3}, \bar{w}_{3})\sigma_{n_e}(w_{4}, \bar{w}_{4})}_\mathcal{M}}\,.\label{TT-dj}
\end{align}
In the above expression, $T_k (w)$ represents the stress energy tensor corresponding to the undeformed $\text{CFT}_2$ living on the $k^{th}$- sheet of the replica manifold $\mathcal{M}_{n_e}$, and $\sigma_{n_e}, \bar{\sigma}_{n_e}$ are the twist operators that are inserted at the endpoints $w_i$ of the intervals \cite{Calabrese:2004eu, Calabrese:2009qy}. Note that in the second line of the above expression, $T^{(n_e)}$ is the energy momentum tensor on the $n_e$-sheeted Riemann surface $\mathcal{M}_{n_e}$ and we have utilized an identity described in \cite{Jeong:2019ylz} to arrive at the second line. To proceed, we transform the energy momentum tensor defined on the cylindrical manifold to the complex plane through \cref{Stress tensor} and apply the following Ward identities \cite{francesco2012conformal},
\begin{align}
	\Braket{T^{(n_e)}(z)\mathcal{O}_{1}(z_{1}, \bar{z}_{1})\ldots \mathcal{O}_{m}(z_{m}, \bar{z}_{m})}_{\mathbb{C}} &= \sum_{j=1}^{m} \left(\frac{h_j}{(z-z_j)^2}+\frac{1}{(z-z_j)}\partial_{z_j}\right) \Braket{\mathcal{O}_{1}(z_{1}, \bar{z}_{1})\ldots \mathcal{O}_{m}(z_{m}, \bar{z}_{m})}_{\mathbb{C}}\,, \notag\\ \Braket{\bar{T}^{(n_e)}(\bar{z})\mathcal{O}_{1}(z_{1}, \bar{z}_{1})\ldots \mathcal{O}_{m}(z_{m}, \bar{z}_{m})}_{\mathbb{C}} &= \sum_{j=1}^{m} \left(\frac{\bar{h}_j}{(\bar{z}-\bar{z}_j)^2}+\frac{1}{(\bar{z}-\bar{z}_j)}\partial_{\bar{z}_j}\right)\Braket{\mathcal{O}_{1}(z_{1}, \bar{z}_{1})\ldots \mathcal{O}_{m}(z_{m}, \bar{z}_{m})}_{\mathbb{C}}\,, \label{ward identity}
\end{align}
where $\mathcal{O}_{i}$ are primary operators with conformal dimensions $(h_i ,\bar{h}_i)$\,.

The expectation value in \cref{TT-dj} may therefore be rewritten as
\begin{align}
	\int_{\mathcal{M}_{n_e}} \Braket{T\bar{T}}_{\mathcal{M}_{n_e}} &= \frac{1}{n_e}\int_{\mathcal{M}}\frac{1}{\Braket{\sigma_{n_e}(z_{1}, \bar{z}_{1})\bar{\sigma}_{n_e}(z_{2}, \bar{z}_{2})\bar{\sigma}_{n_e}(z_{3}, \bar{z}_{3})\sigma_{n_e}(z_{4}, \bar{z}_{4})}_{\mathbb{C}}}\notag\\&\qquad\times\left[-\frac{\pi^2 c\:n_e}{6 \beta^2}+\left(\frac{2 \pi}{\beta} z\right)^2 \sum_{j=1}^{4} \left(\frac{c\left(n_e-\frac{1}{n_e}\right)}{24(z-z_j)^2}+\frac{1}{(z-z_j)}\partial_{z_j}\right)\right]\notag\\&\qquad \times\left[-\frac{\pi^2 c\:n_e}{6 \beta^2}+\left(\frac{2 \pi}{\beta} \bar{z}\right)^2 \sum_{k=1}^{4} \left(\frac{c\left(n_e-\frac{1}{n_e}\right)}{24(\bar{z}-\bar{z}_k)^2}+\frac{1}{(\bar{z}-\bar{z}_k)}\partial_{\bar{z}_k}\right)\right]\notag\\&\qquad\times\Braket{\sigma_{n_e}(z_{1}, \bar{z}_{1})\bar{\sigma}_{n_e}(z_{2}, \bar{z}_{2})\bar{\sigma}_{n_e}(z_{3}, \bar{z}_{3})\sigma_{n_e}(z_{4}, \bar{z}_{4})}_{\mathbb{C}}\,.\label{tt-expect}
\end{align}
In the $t$-channel, where the two disjoint intervals are in proximity, the four point function of twist operators in \cref{tt-expect} is given by \cite{Kulaxizi:2014nma, Malvimat:2017yaj},
\begin{align}
	\Braket{\sigma_{n_e}(z_{1}, \bar{z}_{1})\bar{\sigma}_{n_e}(z_{2}, \bar{z}_{2})\bar{\sigma}_{n_e}(z_{3}, \bar{z}_{3})\sigma_{n_e}(z_{4}, \bar{z}_{4})}_{\mathbb{C}} \approx \left(1-\eta\right)^{h_{n_e}^{(2)}} \left(1-\bar{\eta}\right)^{\bar{h}_{n_e}^{(2)}}\,. \label{4 pt correlator}
\end{align}
The cross ratio $\eta$ is defined as $\eta:=\frac{z_{12} z_{34}}{z_{13} z_{24}}$  with $z_{ij}= (z_i-z_j)$. Now substituting \cref{tt-single sheet} and \cref{tt-expect} into \cref{varEN} and subsequently utilizing \cref{4 pt correlator}, we obtain the first order correction in the entanglement negativity of two disjoint intervals due to the deformation by $\text{T}\bar{\text{T}}$ operator as follows
\begin{align}
	\delta \mathcal{E} (A:B) =& -\frac{\mu c^2 \pi^4}{\beta^4} \int_{\mathcal{M}}\Biggl[-\frac{1}{12} \left(\frac{z^2 z_{12}z_{34}}{(z-z_1)(z-z_2)(z-z_3)(z-z_4)}+\frac{\bar{z}^{2} \bar{z}_{12} \bar{z}_{34}}{(\bar{z}-\bar{z}_{1})(\bar{z}-\bar{z}_{2})(\bar{z}-\bar{z}_{3})(\bar{z}-\bar{z}_{4})}\right)\notag\\&+\frac{1}{4} \left(\frac{z^2 z_{12}z_{34}}{(z-z_1)(z-z_2)(z-z_3)(z-z_4)}\right)\left(\frac{\bar{z}^2 \bar{z}_{12}\bar{z}_{34}}{(\bar{z}-\bar{z}_1)(\bar{z}-\bar{z}_2)(\bar{z}-\bar{z}_3)(\bar{z}-\bar{z}_4)}\right)\Biggr]\,.\label{varENdj-integrand}
\end{align}
\normalsize
The definite integrals in \cref{varENdj-integrand} are evaluated explicitly in \cref{app:disj-int}. Utilizing these results, we obtain 
\small
\begin{align}
	\delta \mathcal{E} (A:B) =& -\frac{\mu c^2 \pi^3}{24\beta^2} \Bigg[ \Bigg(\frac{z_1}{z_{13}}\log \left(\frac{z_1}{z_3}\right)-\frac{z_1}{z_{14}}\log \left(\frac{z_1}{z_4}\right)-\frac{z_2}{z_{23}}\log \left(\frac{z_2}{z_3}\right)+\frac{z_2}{z_{24}}\log \left(\frac{z_2}{z_4}\right)\Bigg)+H.c.\Bigg]+\delta\mathcal{E}_{\text{cross}}
	\notag\\=& -\frac{\mu c^2 \pi^4}{12\beta^3}\Biggl[x_{31}\coth(\frac{\pi x_{31}}{\beta})+x_{42}\coth(\frac{\pi x_{42}}{\beta} )-x_{41}\coth(\frac{\pi x_{41}}{\beta})-x_{32}\coth(\frac{\pi x_{32}}{\beta} )\Biggr]+\delta\mathcal{E}_{\text{cross}}\,. \label{EN-dj}
\end{align}
\normalsize
In the second line we have used $z_i = \bar{z}_i = e^{\frac{2\pi x_i}{\beta}}$ with $\tau_{i}=0$. Note that the additional term $\delta\mathcal{E}_{\text{cross}}$ originates from the crossing correlation of the holomorphic and the anti-holomorphic sector in the integration. 
For two disjoint intervals in proximity, this crossing term is found to be vanishingly small (cf. \cref{app:disj-int}) and hence may be neglected in the leading order.
\subsection{Two adjacent intervals}
\label{subsec:adj-ft}
In this subsection, we will compute $\delta\mathcal{E}(A:B)$ for a bipartite mixed state of two adjacent intervals $A \cup B = [x_1, x_2] \cup [x_2, x_3]$ in a thermal T$\overline{\text{T}}$-deformed CFT$_2$. The computation of the expectation value of the composite operator $\Braket{T\bar{T}}_{\mathcal{M}_{n_e}}$ for the mixed state with two adjacent intervals follows closely from the case of two disjoint intervals:
\begin{align}
	\int_{\mathcal{M}_{n_e}} \Braket{T\bar{T}}_{\mathcal{M}_{n_e}} &= \int_{\mathcal{M}}\frac{1}{n_e} \frac{\Braket{T^{(n_e)}(w)\bar{T}^{(n_e)}(\bar{w})\sigma_{n_e}(w_{1}, \bar{w}_{1})\bar{\sigma}^{2}_{n_e}(w_{2}, \bar{w}_{2})\sigma_{n_e}(w_{3}, \bar{w}_{3})}_{\mathcal{M}}}{\Braket{\sigma_{n_e}(w_{1}, \bar{w}_{1})\bar{\sigma}^{2}_{n_e}(w_{2}, \bar{w}_{2})\sigma_{n_e}(w_{3}, \bar{w}_{3})}_{\mathcal{M}}}\,.\label{TT-adj}
\end{align}
Now, we utilize the conformal transformation specified in \cref{Stress tensor} to transform the energy momentum tensor on the complex plane and employ the Ward identities \cref{ward identity} to obtain,
\begin{align}
	\label{eq:tbartAdj}
	\int_{\mathcal{M}_{n_e}} \Braket{T\bar{T}}_{\mathcal{M}_{n_e}} &= \frac{1}{n_e}\int_{\mathcal{M}}\frac{1}{\Braket{\sigma_{n_e}(z_{1}, \bar{z}_{1})\bar{\sigma}^{2}_{n_e}(z_{2}, \bar{z}_{2})\sigma_{n_e}(z_{3}, \bar{z}_{3})}_{\mathbb{C}}}\notag\\&\qquad\times\bigg[-\frac{\pi^2 c\:n_{e}}{6 \beta^2}+\left(\frac{2 \pi z}{\beta}\right)^2 \sum_{j=1}^{3} \left(\frac{h_j}{(z-z_j)^2}+\frac{1}{(z-z_j)}\partial_{z_j}\right)\bigg]\notag\\&\qquad\times\bigg[-\frac{\pi^2 c\:n_{e}}{6 \beta^2}+\left(\frac{2 \pi \bar{z}}{\beta}\right)^2 \sum_{k=1}^{3} \left(\frac{\bar{h}_k}{(\bar{z}-\bar{z}_k)^2}+\frac{1}{(\bar{z}-\bar{z}_k)}\partial_{\bar{z}_k}\right)\bigg]\notag\\&\qquad\times\Braket{\sigma_{n_e}(z_{1}, \bar{z}_{1})\bar{\sigma}^{2}_{n_e}(z_{2}, \bar{z}_{2})\sigma_{n_e}(z_{3}, \bar{z}_{3})}_{\mathbb{C}}\,.
\end{align}
Here $(h_j, \bar{h}_j)$ refer to the conformal dimensions of twist operator inserted at $(z_j, \bar{z}_j)$. The three point function of twist operators in the above expression is given by \cite{francesco2012conformal},
\begin{align}
	\Braket{\sigma_{n_e}(z_{1}, \bar{z}_{1})\bar{\sigma}^{2}_{n_e}(z_{2}, \bar{z}_{2})\sigma_{n_e}(z_{3}, \bar{z}_{3})}_{\mathbb{C}} = \frac{\mathcal{C}_{\sigma_{n_e}\bar{\sigma}_{n_e}^{2}\sigma_{n_e}}}{z^{h^{(2)}_{n_e}}_{12}z^{h^{(2)}_{n_e}}_{23}z^{2h_{n_e}-h^{(2)}_{n_e}}_{13}\bar{z}^{\bar{h}^{(2)}_{n_e}}_{12}\bar{z}^{\bar{h}^{(2)}_{n_e}}_{23}\bar{z}^{2\bar{h}_{n_e}-\bar{h}^{(2)}_{n_e}}_{13}} \,, \label{3 pt fxn}
\end{align}
where $\mathcal{C}_{\sigma_{n_e}\bar{\sigma}_{n_e}^{2}{\sigma}_{n_e}}$ is the OPE coefficient. Substituting \cref{eq:tbartAdj} and \cref{tt-single sheet} into \cref{varEN}, and utilizing \cref{3 pt fxn}, we obtain the first order correction to the entanglement negativity for two adjacent intervals as follows
\begin{align}
	\delta \mathcal{E}(A:B)=& -\frac{\mu c^2\pi^4}{\beta^4} \int_{\mathcal{M}}\bigg[-\frac{1}{12} \bigg(\frac{z^2 z_{12}z_{23}}{(z-z_1)(z-z_2)^2 (z-z_3)}+\frac{\bar{z}^2 \bar{z}_{12}\bar{z}_{23}}{(\bar{z}-\bar{z}_1)(\bar{z}-\bar{z}_2)^2(\bar{z}-\bar{z}_3)}\bigg) \notag\\&+\frac{1}{4} \bigg(\frac{z^2  z_{12} z_{23}}{(z-z_1)(z-z_2)^2 (z-z_3)}\bigg)\bigg(\frac{\bar{z}^2 \bar{z}_{12} \bar{z}_{23}}{(\bar{z}-\bar{z}_1)(\bar{z}-\bar{z}_2)^2(\bar{z}-\bar{z}_3)}\bigg)\bigg]\,.\label{varEN-adj-integrand}
\end{align}
Once again, we leave the technical details of the integrations in \cref{varEN-adj-integrand} in the \cref{app:adj-int}. The correction to the entanglement negativity is then given by
\begin{align}
	\delta \mathcal{E} (A:B)&= -\frac{\mu c^2\pi^3}{24\beta^2}\bigg[\bigg(\frac{z_1 z_{23}}{z_{12}z_{13}}\log\left(\frac{z_1}{z_2}\right)+\frac{z_{12} z_3}{z_{13} z_{23}}\log \left(\frac{z_2}{z_3}\right)\bigg) + H.c.\bigg]+\delta\mathcal{E}_{\text{cross}}\notag\\&=-\frac{\mu c^2\pi^4}{12\beta^3}\bigg[x_{21}\coth(\frac{\pi x_{21}}{\beta})+x_{32}\coth(\frac{\pi x_{32}}{\beta})-x_{31}\coth(\frac{\pi x_{31}}{\beta})\bigg]+\delta\mathcal{E}_{\text{cross}}\,. \label{EN-adj}
\end{align}
We expect the crossing term $\delta\mathcal{E}_{\text{cross}}$ to vanish similar to the case of two disjoint intervals.

\subsection{A single interval}
\label{subsec:sing-ft}
To compute the entanglement negativity for a single interval $A=[-\ell,0]$ in a $\text{T}\bar{\text{T}}$-deformed CFT$_2$ at finite temperature, we follow the prescription in \cite{Calabrese:2014yza} and introduce two large auxiliary intervals $B_1=[-L, -\ell]$ and $B_2=[0,L]$ sandwiching the interval $A$. The correct entanglement negativity for the mixed state of the single interval $A$ is then obtained by taking the bipartite limit $L\to\infty$ subsequent to the replica limit.

Under the T$\overline{\text{T}}$ deformation, the expectation value of the composite $\text{T}\bar{\text{T}}$ operator for the single interval is obtained as,
\begin{align}
	\int_{\mathcal{M}_{n_e}} \Braket{T\bar{T}}_{\mathcal{M}_{n_e}} &= \int_{\mathcal{M}}\frac{1}{n_e} \frac{\Braket{T^{(n_e)}(w)\bar{T}^{(n_e)}(\bar{w})\sigma_{n_e}(w_{1}, \bar{w}_{1})\bar{\sigma}^{2}_{n_e}(w_{2}, \bar{w}_{2})\sigma^{2}_{n_e}(w_{3}, \bar{w}_{3})\bar{\sigma}_{n_e}(w_{4}, \bar{w}_{4})}}{\Braket{\sigma_{n_e}(w_{1}, \bar{w}_{1})\bar{\sigma}^{2}_{n_e}(w_{2}, \bar{w}_{2})\sigma^{2}_{n_e}(w_{3}, \bar{w}_{3})\bar{\sigma}_{n_e}(w_{4}, \bar{w}_{4})}}\,.\label{TT-sing}
\end{align}
Note that here we have kept the coordinates of the endpoints generic in the correlation function and the specific configuration involving the desired single interval will be considered towards the end of our discussion. Using the transformation of energy momentum tensor from \cref{Stress tensor} and the Ward identities with the energy momentum tensor from \cref{ward identity}, we may simplify \cref{TT-sing} as,
\begin{align}
	\label{eq:tbartsing}
	\int_{\mathcal{M}_{n_e}} \Braket{T\bar{T}}_{\mathcal{M}_{n_e}} =& \int_{\mathcal{M}}\frac{1}{n_e} \frac{1}{\Braket{\sigma_{n_e}(z_{1}, \bar{z}_{1})\bar{\sigma}^{2}_{n_e}(z_{2}, \bar{z}_{2})\sigma^{2}_{n_e}(z_{3}, \bar{z}_{3})\bar{\sigma}_{n_e}(z_{4}, \bar{z}_{4})}}\notag\\&\times\bigg[-\frac{\pi^2 c\: n_{e} }{6 \beta^2}+\left(\frac{2 \pi z}{\beta}\right)^2 \sum_{j=1}^{4} \left(\frac{h_{j}}{(z-z_j)^2}+\frac{1}{(z-z_j)}\partial_{z_j}\right)\bigg]\notag\\&\times\bigg[-\frac{\pi^2 c\: n_{e} }{6 \beta^2}+\left(\frac{2 \pi \bar{z}}{\beta}\right)^2 \sum_{k=1}^{4} \left(\frac{h_k}{(\bar{z}-\bar{z}_k)^2}+\frac{1}{(\bar{z}-\bar{z}_k)}\partial_{\bar{z}_k}\right)\bigg]\notag\\&\times\Braket{\sigma_{n_e}(z_{1}, \bar{z}_{1})\bar{\sigma}^{2}_{n_e}(z_{2}, \bar{z}_{2})\sigma^{2}_{n_e}(z_{3}, \bar{z}_{3})\bar{\sigma}_{n_e}(z_{4}, \bar{z}_{4})}_{\mathcal{C}}\,.
\end{align}
The four point function in the above expression has the following form  \cite{Calabrese:2014yza}, 
\begin{align}
	\Braket{\sigma_{n_e}(z_{1}, \bar{z}_{1})\bar{\sigma}^{2}_{n_e}(z_{2}, \bar{z}_{2})\sigma^{2}_{n_e}(z_{3}, \bar{z}_{3})\bar{\sigma}_{n_e}(z_{4}, \bar{z}_{4})} = c_{n_e}c^{(2)}_{n_e}\left(\frac{\mathcal{F}_{n_e}(\eta)}{z^{2h_{n_e}}_{14} z^{2h^{(2)}_{n_e}}_{23} \eta^{h^{(2)}_{n_e}}}\right)\left(\frac{\bar{\mathcal{F}}_{n_e}(\bar{\eta})}{\bar{z}^{2\bar{h}_{n_e}}_{14} \bar{z}^{2\bar{h}^{(2)}_{n_e}}_{23}\bar{\eta}^{\bar{h}^{(2)}_{n_e}}}\right)\,, \label{correlation-sing}
\end{align}
with the functions $\mathcal{F}_{n_e}(\eta)$ and $\bar{\mathcal{F}}_{n_e}(\bar{\eta})$ obeying the following OPE limits
\begin{align}
	\mathcal{F}_{n_e}(1)\bar{\mathcal{F}}_{n_e}(1)=1 ~~, ~~ \mathcal{F}_{n_e}(0)\bar{\mathcal{F}}_{n_e}(0)=\frac{ \mathcal{C}_{\sigma_{n_e}\bar{\sigma}_{n_e}^{2}\bar{\sigma}_{n_e}}}{c_{n_e}^{(2)}}\,,
\end{align}
where $\mathcal{C}_{\sigma_{n_e}\bar{\sigma}_{n_e}^{2}\bar{\sigma}_{n_e}}$ is the OPE coefficient and $c_{n_e},c_{n_e}^{(2)}$ are normalization constants.
Now, we substitute \cref{eq:tbartsing} and \cref{tt-single sheet} into \cref{varEN} and use the four point function from \cref{correlation-sing} to obtain,
\begin{align}
	\delta \mathcal{E} (A:B)=& -\frac{\mu c^2 \pi^4}{\beta^4} \int_{\mathcal{M}} \Bigg[\frac{1}{12}\Bigg(\frac{z^2}{(z-z_2)^2}+\frac{z^2}{(z-z_3)^2}-\bigg(\sum_{j=1}^{4}\frac{z^2}{(z-z_j)}\partial_{z_j}\bigg)\log\left[z^{2}_{23}\ \eta f(\eta)\right]\Bigg)\notag\\&+\frac{1}{12} \Bigg(\frac{\bar{z}^2}{(\bar{z}-\bar{z}_2)^2}+\frac{\bar{z}^2}{(\bar{z}-\bar{z}_3)^2}-\bigg(\sum_{j=1}^{4}\frac{\bar{z}^2}{(\bar{z}-\bar{z}_j)}\partial_{\bar{z}_j}\bigg)\log\left[\bar{z}^{2}_{23}\bar{\eta}\bar{f}(\bar{\eta})\right]\Bigg)\notag\\&+ \frac{1}{4} \Bigg(\frac{z^2}{(z-z_2)^2}+\frac{z^2}{(z-z_3)^2}-\bigg(\sum_{j=1}^{4}\frac{z^2}{(z-z_j)}\partial_{z_j}\bigg)\log\left[z^{2}_{23}\ \eta f(\eta)\right]\Bigg)\notag\\&\times\Bigg(\frac{\bar{z}^2}{(\bar{z}-\bar{z}_2)^2}+\frac{\bar{z}^2}{(\bar{z}-\bar{z}_3)^2}-\bigg(\sum_{j=1}^{4}\frac{\bar{z}^2}{(\bar{z}-\bar{z}_j)}\partial_{\bar{z}_j}\bigg)\log\left[\bar{z}^{2}_{23}\bar{\eta}\bar{f}(\bar{\eta})\right]\Bigg)\Bigg]\,,\label{varEN-sing-v1}
\end{align}
where we have defined
\begin{align*}
	\lim_{n_e\to 1} \mathcal{F}_{n_e} (\eta) = [f(\eta)]^{c/8} ~~ \text{and} ~~ \lim_{n_e\to 1} \bar{\mathcal{F}}_{n_e} (\bar{\eta}) = [\bar{f}(\bar{\eta})]^{c/8}\,.
\end{align*}
Now we specialize to the specific configuration the single interval of length $\ell$ and subsequently take the bipartite limit $L\rightarrow \infty$. The first order correction in the entanglement negativity of a single interval in a finite temperature CFT$_2$ with T$\overline{\text{T}}$-deformation is therefore given by
\begin{align}
	\delta \mathcal{E} (A:A^c)= -\frac{\mu\pi^4 c^2 \ell}{6\beta^3} \left[-1+\coth(\frac{\pi \ell}{\beta})-e^{-\frac{2\pi \ell}{\beta}}\frac{f'\left(e^{-\frac{2\pi \ell}{\beta}}\right)}{f\left(e^{-\frac{2\pi \ell}{\beta}}\right)}\right]+\delta\mathcal{E}_{\text{cross}}\,.\label{varEN-sing-v2}
\end{align}
The details of integrations to realize \cref{varEN-sing-v2} from \cref{varEN-sing-v1} can be found in \cref{app:sing-int}. We expect the crossing term to vanish in a similar fashion like the earlier cases.  
\section{Holographic entanglement negativity}
\label{sec:HEN}
In this section, we apply the holographic construction\footnote{\label{Ryu-Flam}Note that, an alternative proposal for the holographic entanglement negativity exists in the literature \cite{Kudler-Flam:2018qjo,Kusuki:2019zsp}, which concerns the area of a backreacting bulk cosmic brane homologous to the entanglement wedge cross section (EWCS). In particular for spherical entangling surfaces, the effects of the backreaction from the cosmic brane may be captured by a dimension dependent pre-factor $\chi_d$ and the holographic entanglement negativity is proportional to the area of the minimal EWCS, $\mathcal{E}=\chi_d E_W$. However, it has been demonstrated that the area of the said cosmic brane only agrees with the field theoretic result for the entanglement negativity up to certain constants (possibly related to the holographic \textit{Markov gap} \cite{Hayden:2021gno}), in the context of AdS$_3$/CFT$_2$ \cite{KumarBasak:2021lwm} as well as flat space holography \cite{Basu:2021awn}.} for the entanglement negativity advanced in \cite{Malvimat:2018txq, Jain:2017aqk, Chaturvedi:2016rcn} to the case of various bipartite mixed states in a T$\overline{\text{T}}$-deformed CFT$_2$ defined on a thermal cylinder of circumference $\beta$. As described in \cite{McGough:2016lol}, the holographic dual for a T$\overline{\text{T}}$-deformed CFT$_2$ with deformation parameter $\mu>0$ is given by a portion of AdS$_3$ geometry cut-off at a finite radius $r_C$ with
\begin{align}
	r_C=\sqrt{\frac{6 R^4}{\pi c\mu}}=\frac{R^2}{\epsilon}\,,\label{cut-off radius}
\end{align}
where $R$ is the AdS$_3$ radius, $c$ and $\epsilon$ are the central charge and the UV cut-off of the dual field theory.

Following the proposal in \cite{McGough:2016lol}, the thermal CFT$_2$ with T$\overline{\text{T}}$-deformation is dual to a BTZ black hole \cite{Banados:1998gg} in the finite radius bulk geometry, with the metric
\begin{align}
	\text{d}s^2=\frac{r^2-r_H^2}{R^2}\text{d}t^2+\frac{R^2}{r^2-r_H^2}\text{d}r^2+r^2\text{d}\tilde{x}^2\,,
\end{align}
where $r=r_H$ is the horizon of the black hole. The Euclidean time $t$ is 
identified as $t\sim t+\beta$, where $\beta=\frac{2\pi R^2}{r_H}$ is the inverse temperature of the black hole as well as the dual CFT$_2$. The dual T$\overline{\text{T}}$-deformed CFT$_2$ is located at the cut-off radius $r_C$ and hence the metric of the background manifold is conformal to the flat metric as follows \cite{Chen:2018eqk, Jeong:2019ylz}
\begin{align}
	\text{d}s^2=\text{d}t^2+\frac{\text{d}\tilde{x}^2}{1-\frac{r_H^2}{r_C^2}}\equiv \text{d}t^2+\text{d}x^2\,,
\end{align}
where $x=\tilde{x}\left(1-\frac{r_H^2}{r_C^2}\right)^{-1/2}$ is the spatial coordinate in the CFT$_2$. 

\begin{figure}[h!]
	\centering
	\includegraphics[scale=0.50]{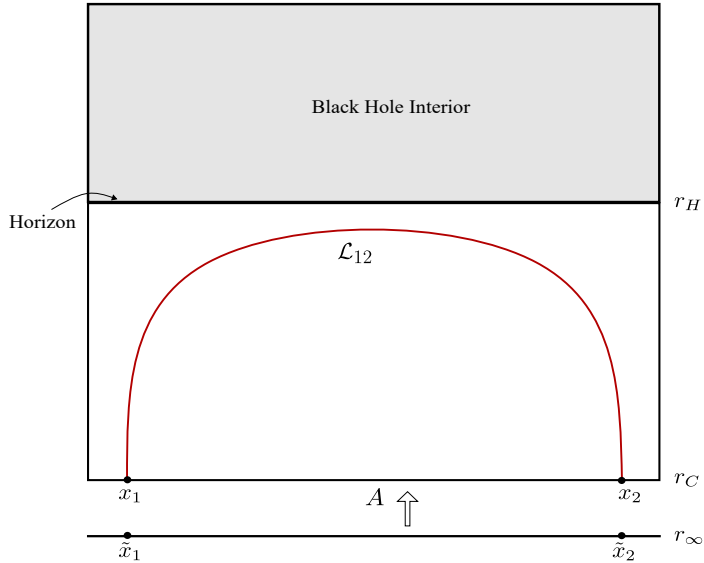}
	\caption{Holographic entanglement entropy for a single interval in a T$\overline{\text{T}}$-deformed CFT$_2$. Figure modified from \cite{Asrat:2020uib}}
	\label{fig:HEE}
\end{figure}

In \cite{Chen:2018eqk,Jeong:2019ylz}, the holographic entanglement entropy for bipartite states in a thermal T$\overline{\text{T}}$-deformed CFT$_2$ was investigated, where it was found that for high temperatures the Ryu-Takayanagi formula \cite{Ryu:2006bv} still applies in the dual finite-radius geometry. The length of the minimal spacelike geodesic homologous to a subsystem $A=[x_i,x_j]$ (cf. \cref{fig:HEE}) in the deformed CFT$_2$ at a temperature $1/\beta$ was computed to be \cite{Chen:2018eqk,Jeong:2019ylz}
\begin{align}
	\mathcal{L}_{ij}=R \log\left(\mathcal{A}(x_i,x_j)+\sqrt{\mathcal{A}(x_i,x_j)^2-1}\right)\,,\label{geod-generic}
\end{align}
where
\begin{align}
	\mathcal{A}(x_i,x_j)\equiv 1+\frac{2 \, r_C^2}{r_H^2}\sinh^2\left(\frac{\pi |x_i-x_j|}{\beta}\sqrt{1-\frac{r_H^2}{r_C^2}}\right)\,.
\end{align}

In the following, we will utilize the above geodesic length to compute the holographic entanglement negativity corresponding to two disjoint, two adjacent and a single interval in a T$\overline{\text{T}}$-deformed CFT$_2$ at finite temperature.
\subsection{Two disjoint intervals}
\label{subsec:disj-hol}
The holographic construction for the entanglement negativity of two disjoint intervals $A$ and $B$ in a CFT$_2$ \cite{Malvimat:2018ood,KumarBasak:2020ams} concerns an algebraic sum of the lengths of bulk minimal spacelike geodesics homologous to various combination of subsystems as follows
\begin{align}
	\mathcal{E}(A:B)&=\frac{3}{16 G_N}\left(\mathcal{L}_{A\cup C}+\mathcal{L}_{B\cup C}-\mathcal{L}_{C}-\mathcal{L}_{A\cup B \cup C}\right)\,,\label{disj-HEN}
\end{align}
where $C$ is another interval sandwiched between $A$ and $B$. Note that the above holographic formula is valid only when the intervals $A$ and $B$ are in close proximity\footnote{On the other hand, for two disjoint intervals which are far away from each other, the holographic entanglement negativity vanishes identically \cite{Malvimat:2018ood,Malvimat:2017yaj}.}.

\begin{figure}[h!]
	\centering
	\includegraphics[scale=0.50]{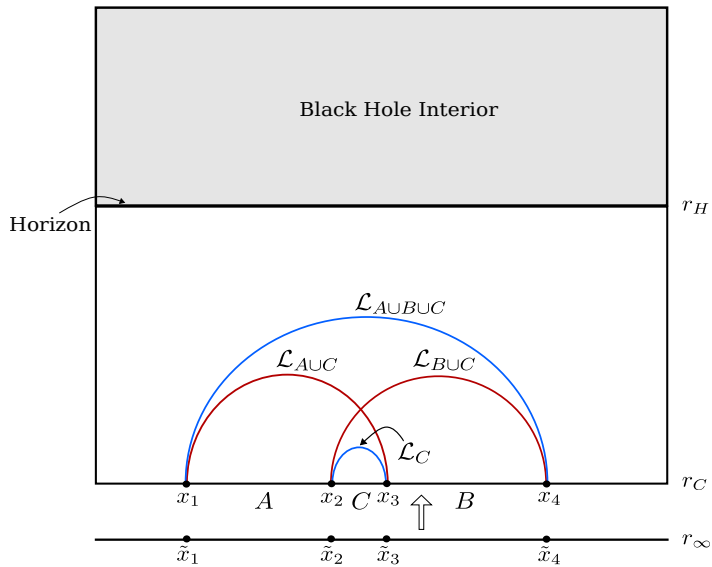}
	\caption{Holographic entanglement negativity for two disjoint intervals in a T$\overline{\text{T}}$-deformed CFT$_2$.}
	\label{fig:HEN-disj}
\end{figure}

Now we apply the above holographic formula \footnote{Note that, the applicability of the formula in \cref{disj-HEN} for a deformed CFT$_2$ is  assumed a priori.} to the case of two disjoint intervals $A=[x_1,x_2]$ and $B=[x_3,x_4]$ in a T$\overline{\text{T}}$-deformed thermal CFT$_2$ defined on a cylinder of circumference $\beta$. The schematic of the setup is depicted in \cref{fig:HEN-disj}. Utilizing \cref{geod-generic} in \cref{disj-HEN}, we obtain
\begin{align}
	\mathcal{E}^{(\mu)}(A:B)&=\frac{3R}{16 G_N}\log\left[\frac{\left(\mathcal{A}(x_1,x_3)+\sqrt{\mathcal{A}(x_1,x_3)^2-1}\right)\left(\mathcal{A}(x_2,x_4)+\sqrt{\mathcal{A}(x_2,x_4)^2-1}\right)}{\left(\mathcal{A}(x_2,x_3)+\sqrt{\mathcal{A}(x_2,x_3)^2-1}\right)\left(\mathcal{A}(x_1,x_4)+\sqrt{\mathcal{A}(x_1,x_4)^2-1}\right)}\right]\label{EN-disj0}\,,
\end{align}
where the superscript indicates a finite deformation parameter for the dual CFT$_2$.
Note that as $r_C\to\infty$, the cut-off radius approaches the asymptotic boundary of the AdS$_3$ geometry and $\mathcal{L}_{ij}$ in \cref{geod-generic} becomes proportional to the holographic entanglement entropy in the usual AdS$_3$/CFT$_2$ setting. In this limit, the above expression reduces to the holographic entanglement negativity for two disjoint intervals in a thermal CFT$_2$, obtained in \cite{Malvimat:2018ood}.

To compare with the field theory computations in \cref{subsec:disj-ft}, we consider the limit of small deformation parameter $\mu$, which corresponds to a large cut-off radius according to \cref{cut-off radius}. Expanding \cref{EN-disj0} for large $r_C$ and further considering the high temperature limit $\beta \ll x_{ij}$ (the dual geometry corresponds to a BTZ black hole only in the high temperature limit), we obtain the entanglement negativity for the disjoint intervals as follows
\begin{align}
	\mathcal{E}^{(\mu)}(A:B)&=\frac{3R}{8G_N}\log\left[\frac{\sinh\left(\frac{\pi x_{13}}{\beta}\right)\sinh\left(\frac{\pi x_{24}}{\beta}\right)}{\sinh\left(\frac{\pi x_{23}}{\beta}\right)\sinh\left(\frac{\pi x_{14}}{\beta}\right)}\right]\notag\\
	\qquad&+\frac{3\pi^4\mu R^2}{16\beta^3 G_N^2}\Bigg[x_{13}\coth\left(\frac{\pi x_{13}}{\beta}\right)+x_{24}\coth\left(\frac{\pi x_{24}}{\beta}\right)-x_{23}\coth\left(\frac{\pi x_{23}}{\beta}\right)-x_{14}\coth\left(\frac{\pi x_{14}}{\beta}\right)\Bigg]\,.
\end{align}
Note that the logarithmic term in the above expression is the holographic entanglement negativity for two disjoint intervals $A=[x_1,x_2]$ and $B=[x_3,x_4]$ in an undeformed holographic CFT$_2$ \cite{Malvimat:2018txq}. On the other hand, the second term proportional to $\mu$ indicates the effects of the T$\overline{\text{T}}$ deformation
which, upon using the holographic dictionary in \cref{cut-off radius} and the usual Brown-Henneaux relation in AdS$_3$/CFT$_2$ \cite{Brown:1986nw}, matches with the field theoretic calculations in \cref{EN-dj} up to the crossing contributions. As discussed earlier, the computations in \cref{app:disj-int} reveal that the crossing integral is vanishingly small in the proximity limit ($\eta\sim 1$). Therefore, the crossing contributions are sub-dominant in the large central charge limit and naturally the holographic computations do not capture their significance.

\subsection{Two adjacent intervals}
\label{subsec:adj-hol}
In the AdS$_3$/CFT$_2$ setup, the entanglement negativity between two adjacent intervals $A$ and $B$ in a holographic CFT$_2$ is proportional to the holographic mutual information as follows \cite{Jain:2017aqk,KumarBasak:2020ams}
\begin{align}
	\mathcal{E}(A:B)&=\frac{3}{16 G_N}\left(\mathcal{L}_{A}+\mathcal{L}_{B}-\mathcal{L}_{A\cup B}\right)\equiv \frac{3}{4}\,\mathcal{I}(A:B)\,,
\end{align}
where in the last equality, the Ryu-Takayanagi formula has been utilized. 

\begin{figure}[h!]
	\centering
	\includegraphics[scale=0.50]{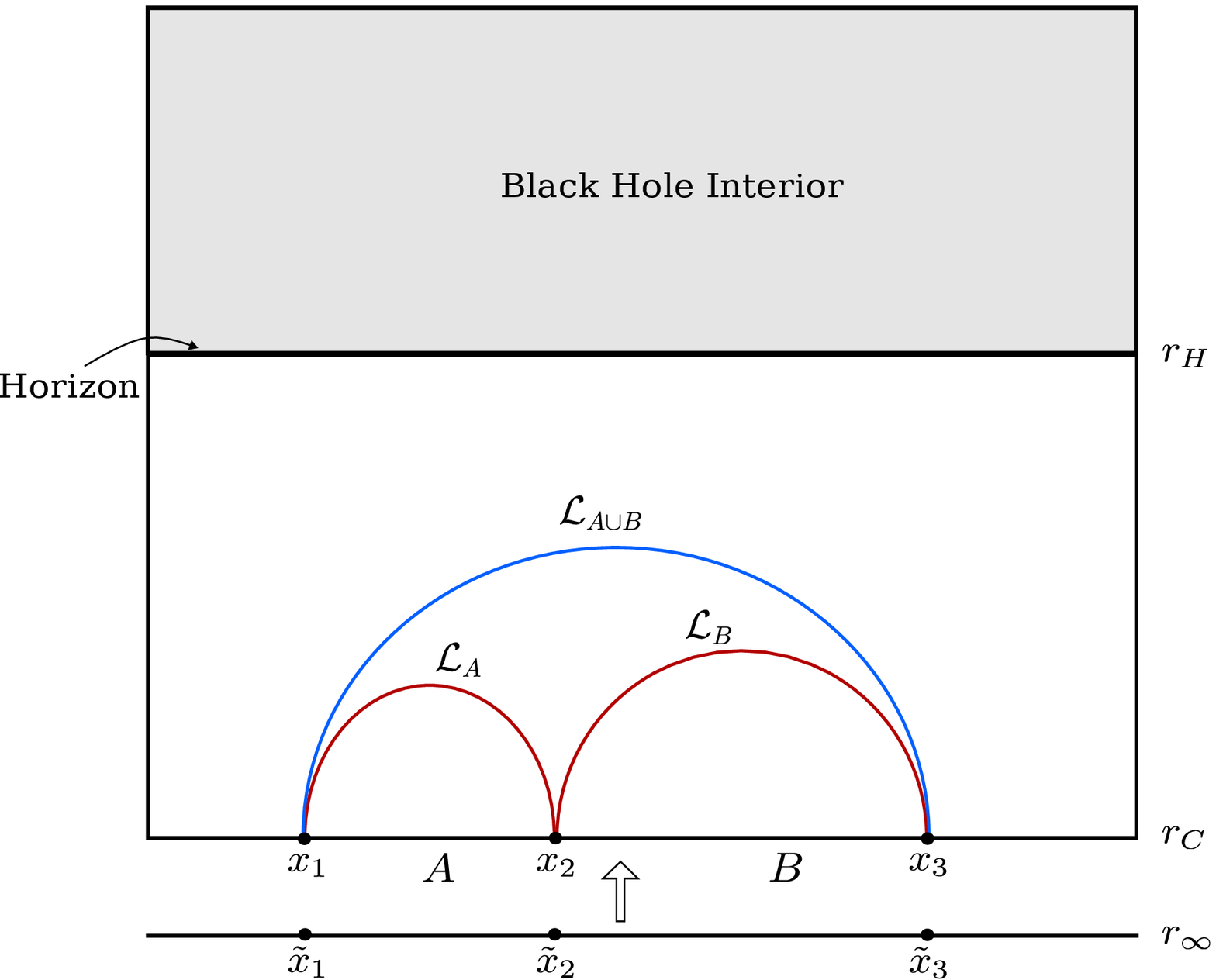}
	\caption{Holographic entanglement negativity for two adjacent intervals in a T$\overline{\text{T}}$-deformed CFT$_2$.}
	\label{fig:HEN-adj}
\end{figure}

For two adjacent intervals $A=[x_1,x_2]$ and $B=[x_2,x_3]$ in a thermal T$\overline{\text{T}}$-deformed CFT$_2$ defined on a temporally compactified cylinder of circumference $\beta$, application of the above holographic formula leads to the expression
\begin{align}
	\mathcal{E}^{(\mu)}(A:B)&=\frac{3R}{16 G_N}\log\left[\frac{\left(\mathcal{A}(x_1,x_2)+\sqrt{\mathcal{A}(x_1,x_2)^2-1}\right)\left(\mathcal{A}(x_2,x_3)+\sqrt{\mathcal{A}(x_2,x_3)^2-1}\right)}{\left(\mathcal{A}(x_1,x_3)+\sqrt{\mathcal{A}(x_1,x_3)^2-1}\right)}\right]\,.\label{EN-adj0}
\end{align}
The schematic of the configuration is sketched in \cref{fig:HEN-adj}.

Expanding the above result for a small deformation of the dual CFT$_2$ at a high temperature $\beta\ll x_{ij}$ and utilizing the holographic dictionary from \cref{cut-off radius}, we obtain
\begin{align}
	\mathcal{E}^{(\mu)}(A:B)&=\frac{3R}{8G_N}\log\left[\frac{\beta}{\pi\epsilon}\frac{\sinh\left(\frac{\pi x_{12}}{\beta}\right)\sinh\left(\frac{\pi x_{23}}{\beta}\right)}{\sinh\left(\frac{\pi x_{13}}{\beta}\right)}\right]\notag\\
	\qquad&+\frac{3\pi^4\mu R^2}{16\beta^3 G_N^2}\Bigg[x_{12}\coth\left(\frac{\pi x_{12}}{\beta}\right)+x_{23}\coth\left(\frac{\pi x_{23}}{\beta}\right)-x_{13}\coth\left(\frac{\pi x_{13}}{\beta}\right)\Bigg]\,.
\end{align}
Once again, the first term on the right hand side corresponds to the holographic entanglement negativity in the usual AdS$_3$/CFT$_2$ scenario \cite{Jain:2017aqk}. The term proportional to $\mu$ corresponds to the leading order corrections due to the deformation of the CFT$_2$
which, upon using the holographic dictionary in \cref{cut-off radius} and the usual Brown-Henneaux relation in AdS$_3$/CFT$_2$ \cite{Brown:1986nw}, matches with the field theoretic calculations in \cref{EN-adj} up to the crossing contributions. Once again, this is expected since the crossing terms are vanishingly small as described in the appendix and therefore do not contribute in the large central charge limit.

\subsection{A single interval}
\label{subsec:sing-hol}
In the context of AdS$_3$/CFT$_2$ correspondence, the holographic characterization of entanglement negativity for a single interval $A$ at a finite temperature requires the introduction of two auxiliary large but finite intervals $B_1\,,\, B_2$ sandwiching the single interval in question \cite{Calabrese:2014yza}. One then computes yet another algebraic sum of the lengths of minimal bulk spacelike geodesics homologous to certain combination of the subsystems involved \cite{Chaturvedi:2016rcn,KumarBasak:2020ams}. Finally, the bipartite limit $B_1\cup B_2\to A^c$ leads to the correct holographic entanglement negativity for $A$ as follows
\begin{align}
	\mathcal{E}(A:A^c)=\lim_{B_1\cup B_2\to A^c} \frac{3}{16 G_N}\left(2\mathcal{L}_A+\mathcal{L}_{B_1}+\mathcal{L}_{B_2}-\mathcal{L}_{A\cup B_1}-\mathcal{L}_{A\cup B_2}\right)\,.
\end{align}

\begin{figure}[h!]
	\centering
	\includegraphics[scale=0.50]{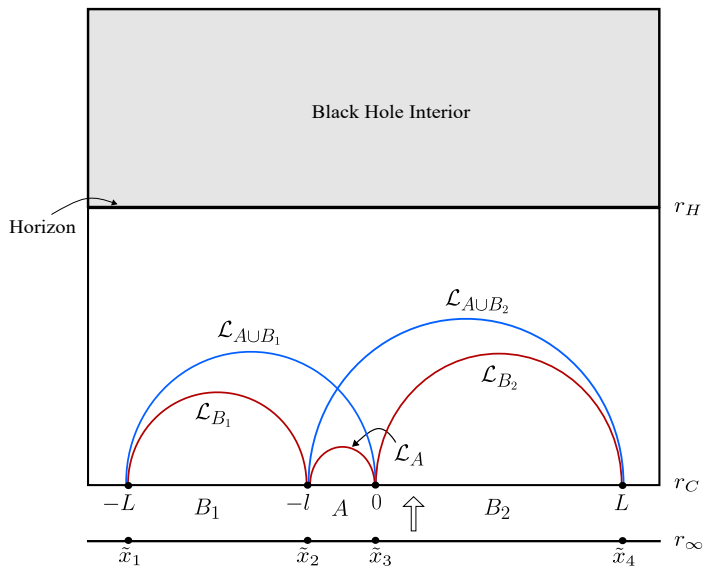}
	\caption{Holographic entanglement negativity for a single interval in a T$\overline{\text{T}}$-deformed CFT$_2$.}
	\label{fig:HEN-sing}
\end{figure}

Now we consider a single interval $A=[-\ell,0]$ in the thermal CFT$_2$ with a T$\overline{\text{T}}$-deformation. Similar to the case of a single interval in an undeformed thermal CFT$_2$ described in \cite{Calabrese:2014yza,Chaturvedi:2016rcn}, we introduce two large auxiliary intervals $B_1=[-L,-\ell]$ and $B_2=[0,L]$ sandwiching the interval $A$ in question. The schematic of the setup is depicted in \cref{fig:HEN-sing}.

Utilizing the length of the minimal boundary anchored spacelike geodesic given in \cref{geod-generic}, the entanglement negativity between $A$ and $B\equiv B_1\cup B_2$ may be obtained as follows
\begin{align}
	\mathcal{E}^{(\mu)}(A:B)&=\frac{3}{4}\Big(\mathcal{I}(A:B_1)+\mathcal{I}(A:B_2)\Big)\notag\\
	&=\frac{3R}{16 G_N}\log\left[\frac{\left(\mathcal{A}(-L,-\ell)+\sqrt{\mathcal{A}(-L,-\ell)^2-1}\right)\left(\mathcal{A}(-\ell,0)+\sqrt{\mathcal{A}(-\ell,0)^2-1}\right)}{\left(\mathcal{A}(-L,0)+\sqrt{\mathcal{A}(-L,0)^2-1}\right)}\right]\notag\\
	&\quad+\frac{3R}{16 G_N}\log\left[\frac{\left(\mathcal{A}(-\ell,0)+\sqrt{\mathcal{A}(-\ell,0)^2-1}\right)\left(\mathcal{A}(0,L)+\sqrt{\mathcal{A}(0,L)^2-1}\right)}{\left(\mathcal{A}(-\ell,L)+\sqrt{\mathcal{A}(-\ell,L)^2-1}\right)}\right]\,.\label{EN-sing0}
\end{align}
In the limit of a large cut-off radius $r_{C}$, the leading order expression for the entanglement negativity at high temperature reduces to
\begin{align}
	\mathcal{E}^{(\mu)}(A:B)&=\frac{3R}{8 G_N}\log\left[\frac{\beta^2}{\pi^2 \epsilon^2} \frac{\sinh^2\left(\frac{\pi \ell}{\beta}\right)\sinh\left(\frac{\pi (L-\ell)}{\beta}\right)}{\sinh\left(\frac{\pi (L+\ell)}{\beta}\right)}\right]\notag\\
	&+\frac{3\mu R^2 \pi^4}{16 G_N^2 \beta^3}\left[(L+\ell)\coth\left(\frac{\pi(L+\ell)}{\beta}\right)-(L-\ell)\coth\left(\frac{\pi(L-\ell)}{\beta}\right)-2\ell\coth\left(\frac{\pi \ell}{\beta}\right)\right]\,.
\end{align}
The bipartite limit may now be achieved by making the auxiliary intervals semi-infinite in length. Therefore, the entanglement negativity for the single interval $A$ in the T$\overline{\text{T}}$-deformed CFT$_2$ is obtained as
\begin{align}
	\mathcal{E}^{(\mu)}(A:A^c)&=\frac{3R}{4 G_N}\left[\log\left(\frac{\beta}{\pi\epsilon}\sinh\left(\frac{\pi \ell}{\beta}\right)\right)-\frac{\pi \ell}{\beta}\right]-\frac{3\mu \pi^4 R^2 \ell}{8 \beta^3 G_N^2}\left[-1+\coth\left(\frac{\pi \ell}{\beta}\right)\right]\,.\label{Neg-sing-final}
\end{align}
Once again, the first term on the left hand side corresponds to the entanglement negativity for a single interval in a finite temperature undeformed CFT$_2$ while the terms proportional to $\mu$ correspond to the leading order corrections due to the T$\overline{\text{T}}$- deformation. Note that, in writing \cref{Neg-sing-final}, we have already made use of the holographic dictionary in \cref{cut-off radius}. The above expression matches with the field theoretic computations in \cref{varEN-sing-v2} apart from the crossing term and the non-universal contributions coming from the arbitrary function $f(\eta)$. This is expected, since the non-universal contributions are generally sub-dominant in the large central charge limit as discussed in \cite{Chaturvedi:2016rcn}. Furthermore, the mismatch of the crossing contributions may be interpreted as follows. As seen from the first equality in \cref{EN-sing0} as well as from \cref{EN-adj0}, the entanglement negativity for the single interval is given by the sum of the individual entanglement negativities for the adjacent subsystems $(A,B_1)$ and $(A,B_2)$ respectively \cite{Chaturvedi:2016rcn}. As argued earlier in \cref{subsec:adj-hol}, for adjacent intervals the effects of the crossing integrals are sub-dominant in the large central charge limit. Hence, in the present case we may also neglect the effects of the crossing terms for a holographic T$\overline{\text{T}}$- deformed CFT$_2$.
\section{Summary and discussions}
In this work, we have studied the entanglement negativity for various bipartite mixed states in a thermal T$\overline{\text{T}}$-deformed CFT$_2$ for a small deformation parameter $\mu$. We have developed a perturbative formula for computing the first order corrections to the entanglement negativity for bipartite states utilizing the replica technique. For a bipartite state $\rho_{AB}$ in a deformed CFT$_2$, our formula involves definite integrals of the expectation value of the T$\overline{\text{T}}$ operator over the replica manifold $\mathcal{M}_{n_e}$ obtained by taking an $n_e$-fold cover of the original manifold where the replica index $n_e$ is an even integer. Utilizing the twist operator formalism, these expectation values may be recast into various correlation functions of twist operators placed at the endpoints of the subsystems $A$ and $B$, including appropriate insertions of stress tensors. Subsequently, we have computed the entanglement negativity for two disjoint, two adjacent and a single interval in a T$\overline{\text{T}}$-deformed CFT$_2$ at a finite temperature. The technical details are collected in \cref{app}. Note that the definite integrals of the stress tensor expectation values may be classified into the holomorphic, anti-holomorphic and the mixing categories. The mixing integrals originate from the crossing correlations between the holomorphic and the anti-holomorphic parts and are in general non-vanishing. However, we have found that for two disjoint intervals in proximity, the mixing terms are negligibly small compared to the other contributions and hence may be neglected altogether. 

Furthermore, we have advanced a holographic construction for computing the entanglement negativity in T$\overline{\text{T}}$-deformed CFT$_2$s with a large central charge and sparse spectrum. The holographic dual of such CFTs with irrelevant deformation is given by AdS$_3$ geometries with a finite cut-off $r_C$. Our holographic constructions for the entanglement negativity for different bipartite states involve algebraic sums of the lengths of minimal spacelike geodesics homologous to the subsystems involved. It is interesting to note that the holographic constructions can deal with arbitrary deformation parameters at any temperature. In the high temperature limit, for a small deformation parameter $\mu$ our holographic results match with the corresponding field theoretic calculations with a large central charge, up to the mixing or crossing contributions. This may be interpreted as the mixing terms becoming very small in the large central charge limit as compared to the holomorphic and anti-holomorphic contributions. This provides a non-trivial consistency check of our holographic constructions. It is important to note that according to a refined version of the holographic constructions \cite{KumarBasak:2020ams} based on \cite{Dong:2021clv}, the entanglement negativity is given in terms of the lengths of bulk \textit{cosmic branes} homologous to various combinations of the subsystems involved. These cosmic branes are tensionful objects for a finite replica parameter $n_e$ and hence back-react on the ambient geometry and onto each other non-trivially. We expect that a closer investigation of these back-reactions may reveal a connection to the holographic origin of the crossing correlations. 

Note that, in an earlier work \cite{Asrat:2020uib}, a low temperature expansion of the leading order corrections to the entanglement wedge cross section was investigated and a mismatch was found with the corresponding field theoretic replica technique results. In this context, we have explored the alternative holographic proposal \cite{Kudler-Flam:2018qjo,Kusuki:2019zsp} based on the entanglement wedge cross section and a preliminary exposition\footnote{A rigorous analysis of the entanglement wedge cross section for finite cut-off geometries dual to T$\bar{\text{T}}$ deformed CFT$_2$s goes beyond the scope of the present work and we leave the same for future explorations.} reveals that the leading order correction in the high-temperature limit conforms to the entanglement negativity obtained in the present work. On the other hand, the zeroth order results still differ from the field theoretic entanglement negativity by certain additive constants proportional to the central charge $c$ (cf. \cref{Ryu-Flam}). In light of these differences as well as the work in \cite{Asrat:2020uib}, we expect that the applicability of the alternative holographic proposal for the entanglement negativity to T$\overline{\text{T}}$-deformed CFT$_2$s requires further investigation.

There are various possible future directions to explore, for example a generalization of our construction to higher dimensions. In particular, for sufficiently symmetric setups in higher dimensions, an investigation of the interactions between different cosmic branes may shed light on the holographic counterpart of the mixing between the holomorphic and anti-holomorphic modes. It will also be interesting to extend our formalism to other entanglement and correlation measures such as the odd entanglement entropy \cite{Tamaoka:2018ned}, the balance partial entanglement \cite{Wen:2021qgx} or the entanglement of purification \cite{Takayanagi:2017knl,Caputa:2018xuf}.
\label{sec:summary}

\appendix
\section{The integrals}\label{app}
In this appendix, we will present the details of integrations of \cref{varENdj-integrand}, \cref{varEN-adj-integrand} and \cref{varEN-sing-v1}. Here the integrations have been performed on the cylindrical manifold $\mathcal{M}$ described by the Euclidean complex coordinates $(x, \tau)$ with appropriate limits.
\subsection{Disjoint intervals}\label{app:disj-int}
We will assume that $z_1<z_2<z_3<z_4$ without any loss of generality. In the following, we will systematically evaluate the holomorphic, anti-holomorphic and the crossing contributions to the definite integral in \cref{varENdj-integrand}.
\subsubsection*{Holomorphic integral}
The holomorphic part of the integration in \cref{varENdj-integrand} is given by
\begin{align}
	&\int_{\mathcal{M}} d^{2}w \frac{z^2(z_1-z_2)(z_3-z_4)}{(z-z_1)(z-z_2)(z-z_3)(z-z_4)}\notag\\&=
	\int_{-\infty}^{\infty} dx \int_{0}^{\beta} d\tau \Biggl[\frac{e^{\frac{4\pi (x+i \tau)}{\beta}}(z_1-z_2)(z_3-z_4)}{(e^{\frac{2\pi(x+i \tau)}{\beta}}-z_1)(e^{\frac{2\pi(x+i \tau)}{\beta}}-z_2)(e^{\frac{2\pi(x+i \tau)}{\beta}}-z_3)(e^{\frac{2\pi(x+i \tau)}{\beta}}-z_4)}\Biggr].
\end{align}
Firstly, we carry out the indefinite integration with respect to $\tau$ and find the primitive function to be,
\begin{align}
	-\frac{i \beta}{2\pi}(z_1-z_2)(z_3-z_4)&\Biggl[\frac{z_1 \log(e^{\frac{2\pi(x+i \tau)}{\beta}}-z_1)}{(z_1-z_2)(z_1-z_3)(z_1-z_4)}-\frac{z_2 \log(e^{\frac{2\pi(x+i \tau)}{\beta}}-z_2)}{(z_1-z_2)(z_2-z_3)(z_2-z_4)}\notag\\&-\frac{z_3 \log(e^{\frac{2\pi(x+i \tau)}{\beta}}-z_3)}{(z_1-z_3)(-z_2+z_3)(z_3-z_4)}-\frac{z_4 \log(e^{\frac{2\pi(x+i \tau)}{\beta}}-z_4)}{(z_1-z_4)(-z_2+z_4)(-z_3+z_4)}\Biggr].
\end{align}
The logarithmic functions require a careful investigation before putting the integration limits, $\tau=0 \: \text{and} \: \tau=\beta$, due to the presence of branch cuts. The contribution due to a branch cut is incorporated through the following identity \cite{Chen:2018eqk,Jeong:2019ylz}:
\begin{align}
	\label{eq:logidentity}
	\log\left(e^{\frac{2\pi(x+i \tau)}{\beta}}-z_j\right) \bigg|_{\tau=0}^{\tau=\beta} = \left\{ \begin{array}{ll}
		2 \pi i, & \textrm{$\quad \text{for}\; e^{\frac{2\pi x}{\beta}} > z_j \Leftrightarrow x > \frac{\beta}{2\pi} \log z_j$}\\
		0, & \textrm{$\quad \text{otherwise}$}
	\end{array} \right.
\end{align}
Therefore, the range of the $x$-integrals get modified for each of the four terms in the integrand as follows,
\begin{align}
	\int_{-\infty}^{\infty} dx\rightarrow \int_{\frac{\beta}{2\pi} \log z_j}^{\infty} dx, \qquad \text{for} \; j=1,2,3,4.
\end{align}
Finally, we integrate over $x$ and insert the limits according to the above prescription to obtain,
\begin{align}
	\int_{\mathcal{M}} d^{2}w \left(\frac{z^2(z_1-z_2)(z_3-z_4)}{(z-z_1)(z-z_2)(z-z_3)(z-z_4)}\right)=-\frac{\beta^2}{2\pi} \Bigg[z_1 \left(\frac{1}{z_{13}}\log \left(\frac{z_1}{z_3}\right)-\frac{1}{z_{14}}\log \left(\frac{z_1}{z_4}\right)\right)\notag\\+z_2 \left(-\frac{1}{z_{23}}\log \left(\frac{z_2}{z_3}\right)+\frac{1}{z_{24}}\log \left(\frac{z_2}{z_4}\right)\right)\Bigg]
\end{align}
We have worked out the integration of the anti-holomorphic part through a similar analysis and have found that the result is same as the holomorphic case. 
\subsubsection*{Crossing integral}
Now we will solve for the integration of the mixing term in \cref{varENdj-integrand} which reads,
\begin{align}
	&\int_{\mathcal{M}} d^{2}w \left(\frac{z^2(z_1-z_2)(z_3-z_4)}{(z-z_1)(z-z_2)(z-z_3)(z-z_4)}\right)\left(\frac{\bar{z}^2(\bar{z}_1-\bar{z}_2)(\bar{z}_3-\bar{z}_4)}{(\bar{z}-\bar{z}_1)(\bar{z}-\bar{z}_2)(\bar{z}-\bar{z}_3)(\bar{z}-\bar{z}_4)}\right)\notag\\&=
	\int_{-\infty}^{\infty} dx \int_{0}^{\beta} d\tau \Bigg[\frac{e^{\frac{4\pi (x+i \tau)}{\beta}}(z_1-z_2)(z_3-z_4)}{(e^{\frac{2\pi(x+i \tau)}{\beta}}-z_1)(e^{\frac{2\pi(x+i \tau)}{\beta}}-z_2)(e^{\frac{2\pi(x+i \tau)}{\beta}}-z_3)(e^{\frac{2\pi(x+i \tau)}{\beta}}-z_4)}\times c.c.\Bigg].
\end{align}
The indefinite integration w.r.t. $\tau$ results in,
\begin{align}
	\frac{i \beta}{2 \pi}e^{\frac{8 \pi x}{\beta }}(\mathcal{A}+\mathcal{B}),
\end{align}
where $\mathcal{A}$ and $\mathcal{B}$ are given by,
\begin{align*}
	&\mathcal{A}=\mathcal{A}_1+\mathcal{A}_2+\mathcal{A}_3+\mathcal{A}_4\notag\\&\mathcal{B}=\mathcal{B}_1+\mathcal{B}_2+\mathcal{B}_3+\mathcal{B}_4  
\end{align*}
\begin{align}
	&\mathcal{A}_1=-\frac{z_1^3 (z_3-z_4)(\bar{z}_1-\bar{z}_2)(\bar{z}_3-\bar{z}_4) }{(z_1-z_3) (z_1-z_4) \left(z_1 \bar{z}_1-e^{\frac{4 \pi  x}{\beta}}\right) \left(z_1 \bar{z}_2-e^{\frac{4 \pi  x}{\beta }}\right) \left(z_1 \bar{z}_3-e^{\frac{4 \pi  x}{\beta }}\right) \left(z_1 \bar{z}_4-e^{\frac{4 \pi  x}{\beta }}\right)}\log\left(e^{\frac{2 \pi  (x+i \tau )}{\beta }}-z_1\right),
	\notag\\&\mathcal{A}_2=\frac{ z_2^3 (z_3-z_4)(\bar{z}_1-\bar{z}_2)(\bar{z}_3-\bar{z}_4) }{(z_2-z_3)(z_2-z_4) \left(z_2 \bar{z}_1-e^{\frac{4 \pi  x}{\beta }}\right) \left(z_2 \bar{z}_2-e^{\frac{4 \pi  x}{\beta }}\right) \left(z_2 \bar{z}_3-e^{\frac{4 \pi  x}{\beta }}\right) \left(z_2 \bar{z}_4-e^{\frac{4 \pi  x}{\beta }}\right)}\log \left(e^{\frac{2 \pi  (x+i \tau )}{\beta }}-z_2\right),
	\notag\\&\mathcal{A}_3=\frac{ z_3^3 (z_1-z_2)(\bar{z}_1-\bar{z}_2)(\bar{z}_3-\bar{z}_4)}{(z_1-z_3) (z_3-z_2) \left(z_3 \bar{z}_1-e^{\frac{4 \pi  x}{\beta }}\right) \left(z_3 \bar{z}_2-e^{\frac{4 \pi  x}{\beta }}\right) \left(z_3 \bar{z}_3-e^{\frac{4 \pi  x}{\beta }}\right) \left(z_3 \bar{z}_4-e^{\frac{4 \pi  x}{\beta }}\right)}\log \left(e^{\frac{2 \pi  (x+i \tau )}{\beta }}-z_3\right),
	\notag\\&\mathcal{A}_4=-\frac{ z_4^3 (z_1-z_2)(\bar{z}_1-\bar{z}_2)(\bar{z}_3-\bar{z}_4)}{(z_1-z_4) (z_4-z_2) \left(z_4 \bar{z}_1-e^{\frac{4 \pi  x}{\beta }}\right) \left(z_4 \bar{z}_2-e^{\frac{4 \pi  x}{\beta }}\right) \left(z_4 \bar{z}_3-e^{\frac{4 \pi  x}{\beta }}\right) \left(z_4 \bar{z}_4-e^{\frac{4 \pi  x}{\beta }}\right)}\log \left(e^{\frac{2 \pi  (x+i \tau )}{\beta }}-z_4\right),
	\notag\\&\mathcal{B}_1=\frac{ \bar{z}_1^3 (z_1-z_2)(z_3-z_4)(\bar{z}_3-\bar{z}_4)}{\left(e^{\frac{4 \pi  x}{\beta }}-z_1 \bar{z}_1\right) \left(e^{\frac{4 \pi  x}{\beta }}-z_2 \bar{z}_1\right) \left(e^{\frac{4 \pi  x}{\beta }}-z_3 \bar{z}_1\right) \left(e^{\frac{4 \pi  x}{\beta }}-z_4 \bar{z}_1\right)(\bar{z}_1-\bar{z}_3) (\bar{z}_1-\bar{z}_4)}\log \left(e^{\frac{2 \pi  x}{\beta }}-e^{\frac{2 i \pi  \tau }{\beta }} \bar{z}_1\right),
	\notag\\&\mathcal{B}_2=-\frac{\bar{z}_2^3 (z_1-z_2)(z_3-z_4)(\bar{z}_3-\bar{z}_4)}{(\bar{z}_2-\bar{z}_3)(\bar{z}_2-\bar{z}_4)\left(e^{\frac{4 \pi  x}{\beta }}-z_1 \bar{z}_2\right) \left(e^{\frac{4 \pi  x}{\beta }}-z_2 \bar{z}_2\right) \left(e^{\frac{4 \pi  x}{\beta }}-z_3 \bar{z}_2\right) \left(e^{\frac{4 \pi  x}{\beta }}-z_4 \bar{z}_2\right)}\log \left(e^{\frac{2 \pi  x}{\beta }}-e^{\frac{2 i \pi  \tau }{\beta }} \bar{z}_2\right),\notag\\
	&\mathcal{B}_3=-\frac{\bar{z}_3^3 (z_1-z_2)(z_3-z_4)(\bar{z}_1-\bar{z}_2)}{(\bar{z}_1-\bar{z}_3) (\bar{z}_3-\bar{z}_2) \left(e^{\frac{4 \pi  x}{\beta }}-z_1 \bar{z}_3\right) \left(e^{\frac{4 \pi  x}{\beta }}-z_2 \bar{z}_3\right) \left(e^{\frac{4 \pi  x}{\beta }}-z_3 \bar{z}_3\right) \left(e^{\frac{4 \pi  x}{\beta }}-z_4 \bar{z}_3\right)} \log \left(e^{\frac{2 \pi  x}{\beta }}-e^{\frac{2 i \pi  \tau }{\beta }} \bar{z}_3\right),
	\notag\\&\mathcal{B}_4=\frac{\bar{z}_4^3(z_1-z_2)(z_3-z_4)(\bar{z}_1-\bar{z}_2) }{ (\bar{z}_1-\bar{z}_4) (\bar{z}_4-\bar{z}_2) \left(e^{\frac{4 \pi  x}{\beta }}-z_1 \bar{z}_4\right) \left(e^{\frac{4 \pi  x}{\beta }}-z_2 \bar{z}_4\right) \left(e^{\frac{4 \pi  x}{\beta }}-z_3 \bar{z}_4\right) \left(e^{\frac{4 \pi  x}{\beta }}-z_4 \bar{z}_4\right)}\log \left(e^{\frac{2 \pi  x}{\beta }}-e^{\frac{2 i \pi  \tau }{\beta }} \bar{z}_4\right).
\end{align}
The identity of \cref{eq:logidentity} suggests the following modifications of the $x$- integration limits,
\begin{align*}
	\mathcal{A}_k:\int_{-\infty}^{\infty} dx \rightarrow \int_{\frac{\beta}{2\pi}\log z_k}^{\infty} dx\,\,,  \qquad\mathcal{B}_k:\int_{-\infty}^{\infty} dx \rightarrow \int_{-\infty}^{\frac{\beta}{2\pi}\log \bar{z}_k} dx.
\end{align*}
After integrating over $x$ with the above limits, we finally determine the crossing integral as,
\begin{align}
	&\frac{\beta^2}{2\pi |z_{13}|^2 |z_{23}|^2 |z_{14}|^2 |z_{24}|^2}\notag\\ &\times\Big[-z_1z_{23}z_{24}z_{34}\Big(\bar{z}_1\bar{z}_{23}\bar{z}_{24}\bar{z}_{34}\log z_{1\bar{1}}-\bar{z}_2\bar{z}_{13}\bar{z}_{14}\bar{z}_{34}\log z_{1\bar{2}}+\bar{z}_3\bar{z}_{12}\bar{z}_{14}\bar{z}_{24}\log z_{1\bar{3}}-\bar{z}_4\bar{z}_{12}\bar{z}_{13}\bar{z}_{23} \log z_{1\bar{4}}\Big)\notag\\&+ z_2 z_{13} z_{14}z_{34}\Big(\bar{z}_1\bar{z}_{23}\bar{z}_{24}\bar{z}_{34}\log z_{2\bar{1}}-\bar{z}_2\bar{z}_{13}\bar{z}_{14}\bar{z}_{34}\log z_{2\bar{2}}+\bar{z}_3\bar{z}_{12}\bar{z}_{14}\bar{z}_{24}\log z_{2\bar{3}}-\bar{z}_4\bar{z}_{12}\bar{z}_{13}\bar{z}_{23} \log z_{2\bar{4}}\Big)\notag\\&-z_3 z_{12} z_{14} z_{24}\Big(\bar{z}_1\bar{z}_{23}\bar{z}_{24}\bar{z}_{34}\log z_{3\bar{1}}-\bar{z}_2\bar{z}_{13}\bar{z}_{14}\bar{z}_{34}\log z_{3\bar{3}}+\bar{z}_3\bar{z}_{12}\bar{z}_{14}\bar{z}_{24}\log z_{3\bar{3}}-\bar{z}_4\bar{z}_{12}\bar{z}_{13}\bar{z}_{23} \log z_{3\bar{4}}\Big)\notag\\&+ z_4 z_{12} z_{13} z_{23}\Big(\bar{z}_1\bar{z}_{23}\bar{z}_{24}\bar{z}_{34}\log z_{4\bar{1}}-\bar{z}_2\bar{z}_{13}\bar{z}_{14}\bar{z}_{34}\log z_{4\bar{2}}+\bar{z}_3\bar{z}_{12}\bar{z}_{14}\bar{z}_{24}\log z_{4\bar{3}}-\bar{z}_4\bar{z}_{12}\bar{z}_{13}\bar{z}_{23} \log z_{4\bar{4}}\Big)\Big].\label{disj-crossing-final}
\end{align}
We have explicitly checked that the expression in \cref{disj-crossing-final} vanishes when the disjoint intervals are in proximity to each other.
\subsection{Adjacent intervals}\label{app:adj-int}
The holomorphic part of the integral in \cref{varEN-adj-integrand} is given by
\begin{align}
	\int_{\mathcal{M}} d^2 w \frac{z^2 (z_1-z_2)(z_2-z_3)}{(z-z_1)(z-z_2)^2 (z-z_3)}=\int_{-\infty}^{\infty} dx \int_{0}^{\beta} d\tau \frac{e^{\frac{4\pi(x+i\tau)}{\beta}}(z_1-z_2)(z_2-z_3)}{\left(e^{\frac{2\pi(x+i\tau)}{\beta}}-z_1\right)\left(e^{\frac{2\pi(x+i\tau)}{\beta}}-z_2\right)^2 \left(e^{\frac{2\pi(x+i\tau)}{\beta}}-z_3\right)}.
\end{align}
We use a similar procedure as in \cref{app:disj-int}. First we integrate over $\tau$ and obtain the primitive function to be
\begin{align}
	\frac{i\beta}{2\pi} z_{12} z_{23} &\Bigg[-\frac{z_2}{z_{12} z_{23}\Big(e^{\frac{2\pi(x+i\tau)}{\beta}}-z_2\Big)}-\frac{z_1 \log \left(e^{\frac{2\pi(x+i\tau)}{\beta}}-z_1\right)}{z_{12}^2 z_{13}}\notag\\&+\frac{\left(z_{2}^2-z_1 z_3\right) \log \left(e^{\frac{2\pi(x+i\tau)}{\beta}}-z_2\right)}{z_{12}^2 z_{23}^2}+\frac{z_3 \log \Big(e^{\frac{2\pi(x+i\tau)}{\beta}}-z_3\Big)}{z_{13} z_{23}^2}\Bigg].
\end{align}
The first term in the above expression vanishes when we insert the integration limits $\tau = 0$ and $\tau = \beta$, whereas the logarithmic function contributes according to the identity in \cref{eq:logidentity}. Again, the branch cut of the logarithmic function changes the limits of integration over $x$ as follows, 
\begin{align*}
	\int_{-\infty}^{\infty} dx \rightarrow \int_{\frac{\beta}{2\pi} \log z_j}^{\infty} dx, \quad \text{for}\: j=1, 2, 3.
\end{align*}
Finally, we perform the $x$-integration to obtain,
\begin{align}
	\int_{\mathcal{M}} d^2 w \frac{z^2 z_{12} z_{23}}{(z-z_1)(z-z_2)^2 (z-z_3)}=-\frac{\beta^2}{2\pi z_{12}z_{13}z_{23}}\Bigg[z_1 z_{23}^2 \log\left(\frac{z_1}{z_2}\right)+z_{12}^2 z_3\log\left(\frac{z_2}{z_3}\right)\Bigg].\label{adjint}
\end{align}

The anti-holomorphic integral may also be tackled in a similar fashion. For our case with spatial intervals $(z_i \in \mathbb{R})$, we find that integration over the anti-holomorphic part exactly reproduces \cref{adjint}.
We can also evaluate the crossing integral for two adjacent intervals from first principles utilizing the method outlined in \cref{app:disj-int}. Alternatively, we may take the adjacent limit $z_3\to z_2\,,\bar{z}_3\to \bar{z}_2$ of the disjoint crossing integral in \cref{disj-crossing-final} to write
\begin{align*}
	\delta\mathcal{E}^{(\text{adj})}_{\text{cross}}=\delta\mathcal{E}^{(\text{disj})}_{\text{cross}}\left(z_3\to z_2\,,\bar{z}_3\to \bar{z}_2\right)\,.
\end{align*}

\subsection{Single interval}\label{app:sing-int}
The holomorphic part of the integration in \cref{varEN-sing-v1} is,
\begin{align}
	&\int_{\mathcal{M}} d^2 w \bigg[\frac{z^2}{(z-z_2)^2}+\frac{z^2}{(z-z_3)^2}-\sum_{j=1}^{4}\frac{z^2}{(z-z_j)}\partial_{z_j}\log\left[z^{2}_{23}\ \eta f(\eta)\right]\bigg]\notag\\&=-\int_{-\infty}^{\infty} dx \int_{0}^{\beta} d\tau \frac{e^{\frac{4\pi(x+i\tau)}{\beta}} z_{23}}{\left(e^{\frac{2\pi(x+i\tau)}{\beta}}-z_1\right)\left(e^{\frac{2\pi(x+i\tau)}{\beta}}-z_2\right)^2 \left(e^{\frac{2\pi(x+i\tau)}{\beta}}-z_3\right)^2 \left(e^{\frac{2\pi(x+i\tau)}{\beta}}-z_4\right)} \notag\\&~~\times\bigg[z_1 z_2 z_{34}+z_{12} z_3 z_4 +(z_{12} - z_{34})e^{\frac{4\pi(x+i\tau)}{\beta}} +2(z_1 z_3 -z_2 z_4)e^{\frac{2\pi(x+i\tau)}{\beta}} \notag\\
	&\qquad\qquad\qquad\qquad\qquad\qquad\qquad\qquad\qquad\quad~+\left(e^{\frac{2\pi(x+i\tau)}{\beta}}-z_2\right)\left(e^{\frac{2\pi(x+i\tau)}{\beta}}-z_3\right)z_{14} \frac{\eta\,f'\left(\eta\right)}{f \left(\eta\right)}\bigg].
\end{align}
Firstly, we perform an indefinite integration over $\tau$ and obtain,
\begin{align}
	-\frac{i \beta}{2\pi} \frac{z_{23}}{z_{13} z_{24} \;f\left(\eta\right)} (\mathcal{C}+\mathcal{D}), \label{singtau}
\end{align}
where $\mathcal{C}$ and $\mathcal{D}$ are given as,
\begin{align}
	&\mathcal{C}=-\frac{z_{13} z_{24}}{z_{23}} f \left(\eta\right) \left[\frac{z_2}{\left(e^{\frac{2\pi (x+i\tau)}{\beta}}-z_2\right)}+\frac{z_3}{\left(e^{\frac{2\pi (x+i\tau)}{\beta}}-z_3\right)}\right],\notag\\
	&\mathcal{D}= \Bigg[f \left(\eta\right)+\;\eta\; f' \left(\eta\right)\Bigg] (\mathcal{D}_1+\mathcal{D}_4)+\frac{1}{z_{23}^2} (\mathcal{D}_2+\mathcal{D}_3),\notag\\
	&\mathcal{D}_1= -\frac{z_1 z_{24}}{z_{12}} \log(e^{\frac{2\pi (x+i\tau)}{\beta}}-z_1),\notag\\
	&\mathcal{D}_2= \frac{z_{13}}{z_{12}}\bigg[\Big[z_2(z_2^{2}+z_2(z_3-2z_4)-z_1 (2z_3-z_4))+z_1 z_3 z_4 \Big] f \left(\eta\right)+z_2 z_{23}z_{14} \;\eta\; f' \left(\eta\right)\bigg] \log(e^{\frac{2\pi (x+i\tau)}{\beta}}-z_2),\notag\\
	&\mathcal{D}_3= -\frac{z_{24}}{z_{34}}\bigg[\Big[z_3(z_3^{2}+ z_2(z_3-2z_4)-z_1(2 z_3- z_4))+z_1 z_2 z_4\Big]f \left(\eta\right)+z_3 z_{23} z_{14}\;\eta\; f' \left(\eta\right)\bigg]\log(e^{\frac{2\pi (x+i\tau)}{\beta}}-z_3),\notag\\
	&\mathcal{D}_4= \frac{z_4 z_{13}}{z_{34}}\log(e^{\frac{2\pi (x+i\tau)}{\beta}}-z_4).
\end{align}
We notice that the expression $\mathcal{C}$ vanishes when we insert the limits of integration $\tau = 0$ and $\tau= \beta$, whereas the logarithmic terms of $\mathcal{D}$ contribute through the identity described in \cref{eq:logidentity}. Therefore, the limits of the  $x$-integration get modified as follows,
\begin{align}
	\mathcal{D}_k:\int_{-\infty}^{\infty} dx \rightarrow \int_{\frac{\beta}{2\pi}\log z_k}^{\infty} dx, \quad k=1, 2, 3, 4. \label{singintlimit}
\end{align}
Subsequently, performing the $x$-integration over \cref{singtau}, considering the modified integration limits from \cref{singintlimit}, we obtain 
\begin{align}
	\frac{\beta^2}{2\pi z_{13}^2 z_{24}^2} \Bigg[\frac{\eta}{z_{23}} &\bigg(z_1 z_{23}^2 z_{24} z_{34} \log z_1- z_{13} z_{34} (z_2 (z_2^{2}+z_{2}(z_3-2z_4)-z_1(2z_3-z_4))+z_1 z_3 z_4)\log z_2\notag\\&+z_{12}z_{24}(z_3(z_3^{2}+z_2 (z_3-2 z_4)-z_1 (2z_3-z_4))+z_1 z_2 z_4)\log z_3-z_{12} z_{13} z_{23}^2 z_4 \log z_4 \bigg)\notag\\&+\bigg(z_1 z_{23} z_{24} z_{34}\log z_1 - z_{2} z_{13} z_{14} z_{34} \log z_{2} + z_{12} z_{14} z_{24} z_3 \log z_{3}- z_{12} z_{13} z_{23} z_4 \log z_{4}\bigg)\frac{f' \left(\eta\right)}{f \left(\eta\right)}\Bigg].
\end{align}
Finally, we consider the specific case of a single interval of length $\ell$ via the substitutions $\{z_1, z_2, z_3, z_4\} \rightarrow \{e^{-\frac{2\pi L}{\beta}}, e^{-\frac{2\pi \ell}{\beta}}, 1, e^{\frac{2\pi L}{\beta}}\}$ and subsequently take the bipartite limit to obtain
\begin{align}
	&\lim_{L\to\infty}\int_{\mathcal{M}} d^2 w \left[\frac{z^2}{(z-z_2)^2}+\frac{z^2}{(z-z_3)^2}-\sum_{j=1}^{4}\frac{z^2}{(z-z_j)}\partial_{z_j}\log\left[z^{2}_{23}\ \eta f(\eta)\right]\right]\notag\\&=
	\ell \beta \left[-1+\coth \left(\frac{\pi \ell}{\beta}\right)-e^{-\frac{2\pi \ell}{\beta}}\frac{f'\left(e^{-\frac{2\pi \ell}{\beta}}\right)}{f\left(e^{-\frac{2\pi \ell}{\beta}}\right)}\right].
\end{align} 
It may be shown by a similar procedure that the antiholomorphic integral also gives the same result which follows from our consideration of a spatial interval of length $\ell$ on the cylinder. In this case also we expect the crossing term to be vanishingly small as argued in \cref{subsec:sing-hol}.
	\bibliographystyle{utphys}
	\bibliography{reference}
	
\end{document}